\documentclass[11pt]{article}
\usepackage{amsmath,times,mathptm,amssymb,subfigure}
\usepackage{graphicx,version}

\numberwithin{equation}{section}
\def\be{\begin{eqnarray}}
\def\ee{\end{eqnarray}}
\def\benl{\begin{eqnarray*}}
\def\eenl{\end{eqnarray*}}

\usepackage{epsfig}
\usepackage{amsfonts}

\def\beq{\begin{equation}}
\def\eeq{\end{equation}}

\def\part#1{\frac{\partial #1}{\partial t}}

\newcommand{\ben}{\begin{eqnarray}}
\newcommand{\een}{\end{eqnarray}}
\newcommand{\bea}{\begin{array}}
\newcommand{\eea}{\end{array}}
\newcommand{\bef}{\begin{figure}}
\newcommand{\eef}{\end{figure}}
\begin{document}


\title{\bf  Quasi-incompressible Multi-species Ionic Fluid Models}
\author{
{\bf Xiaogang Yang}\footnote{School of Science, Wuhan Institute of Technology, Wuhan City, Hubei Province, P. R. China, 430205. Email: xgyang@wit.edu.cn.},
{\bf Yuezheng Gong}\footnote{College of Science, Nanjing University of Aeronautics and Astronautics, Nanjing 210016, P. R.
China. Email: gongyuezheng@nuaa.edu.cn.},
{\bf Jun Li}\footnote{School of Mathematics, Tianjin Normal University, Tianjin, P. R. China, 300071. Email: nkjunli@gmail.com.},
{\bf Robert S. Eisenberg}\footnote{ Department of Molecular Biophysics \& Physiology, Rush University Medical Center, Chicago, IL 60612. Email: bob.eisenberg@gmail.com.}
and
{\bf Qi Wang}\footnote{Corresponding author: Department of Mathematics,  University of South Carolina, Columbia, SC 29028, USA;  Beijing Computational Science Research Center, Beijing, P. R. China, 100193. Email: qwang@math.sc.edu.}}
\date{}
\maketitle
\begin{abstract}
In traditional hydrodynamic theories for ionic fluids, conservation of the mass and linear momentum is not properly taken care of. In this paper, we develop  hydrodynamic theories for a viscous, ionic fluid of $N$ ionic species enforcing mass and momentum conservation as well as considering the size effect of the ionic particles.   The  theories developed are quasi-incompressible in that the mass-average velocity is no longer divergence-free whenever there exists variability in densities of  the fluid components, and the models are dissipative. We present several ways to derive the transport equations for the ions, which lead to different rates of energy dissipation. The theories can be formulated in either number densities, volume fractions or mass densities  of the ionic fluid components. We show that the theory with the Cahn-Hilliard transport equation for ionic species reduces to the classical Poisson-Nernst-Planck (PNP) model with the size effect for ionic fluids when the densities of the fluid components are equal and the entropy of the solvent is neglected. It further reduces to the PNP model when the size effect is neglected.
A linear stability analysis of the model together with two of its limits, which is the extended PNP model (EPNP defined in the text) and the classical PNP model (CPNP) with the finite size effect, on a constant state and a comparison among the three models in 1D space are presented to highlight the similarity and the departure of this model from the EPNP and the CPNP model.
\end{abstract}
\vskip 12 pt


\section{Introduction}

\noindent \indent
Phase field models have been used successfully to study a variety of multiphasic  phenomena like equilibrium shapes of vesicle membranes  \cite{DuLiRyWa05,DuLiWa04}, blends of polymeric liquids \cite{Wang&E&L&Z2002,Wa02,Wang&F&Z2004,Forest&W2005}, multiphase fluid flows \cite{FeLiShYu05,Lin&C&W2007,LiWa01,LoTr98,LiSh02,WiLoKiJo04,Wheeler&M&B1996,YaFeLiSh06,YuFeLiSh04,Zhang&C&W2008}, dentritic growth in solidification, microstructure evolution \cite{Kobayashi1993,McFadden&W&B&C&S1993,Karma&R1999}, grain growth \cite{Chen&Y1994},  crack propagation \cite{Chen2002}, morphological pattern formation in thin films and on surfaces \cite{Li&H&L&C2001,SeHuLiShOhCh03},  self-assembly dynamics of two-phase monolayers on an elastic substrate   \cite{Lu&S2001},
a wide variety of diffusive and diffusion-less solid-state phase transitions \cite{Chen&W1996,Wang&C1999},
dislocation modeling in microstructure, electro-migration and multiscale modeling \cite{Tadmor&P&O1996}. Multiple phase-field methods can be devised to study multiphase materials \cite{Wheeler&M&B1996}.
Recently, phase field models are applied to study   liquid crystal drop deformation in another fluid, liquid films, polymer nanocomposites,  biofilms and cells \cite{FeLiShYu05,Lin&C&W2007,LiWa01,LoTr98,LiSh02,WiLoKiJo04,Wheeler&M&B1996,YaFeLiSh06,YuFeLiSh04,
YuFeLiSh05,Forest&L&W2009,Zhang&W2009,Brandon&W2010,Zhao&W2015,Zhao&W2016-1,Zhao&W2016-2}.

Comparing to other mathematical and computational technologies available for studying multi-phase materials, the phase-field approach exhibits a clear advantage in its simplicity in model formulation, ease of numerical implementation, and the ability to explore essential interfacial physics at the interfacial regions etc. Computing the interface without explicitly tracking the interface is the most attractive numerical feature of this modeling and computational technology. Since the pioneering work of Cahn and Hilliard in the 50's of the last century, the Cahn-Hilliard equation has been the foundation for various phase field models
\cite{Cahn&H1958,Cahn&H1959}. It arises naturally as a model for multiphase material mixtures should the entropic and mixing energy of the mixture system be known.

While modeling immiscible binary fluid mixtures using phase field theories, one commonly uses a labeling or a phase variable (also known as a volume fraction or an order parameter) $\phi$ to distinguish between distinct fluid phases. For instance $\phi=1$ indicates one fluid phase while $\phi=0$ denotes the other fluid phase in an immiscible binary mixture. The interfacial region is tracked by $0<\phi<1$. For historical more than logical reasons, most mixing energies are calculated in terms of the volume fraction instead of the mass fraction in the literature \cite{Flory1953,Doi&See}. Consequently, the system free energy including the entropic and mixing contribution has been formulated in terms of the volume fraction as well \cite{Flory1953,Doi&See}, given in the form  $F[\phi, \nabla \phi,\cdots]$. A transport equation for the volume fraction $\phi$ along with the conservation equation of momentum and the continuity equation constitute the essential part of the governing system of hydrodynamic equations for the binary fluid mixture, where the volume fraction serves as an internal variable for the fluid mixture.

In this formulation, the material incompressibility is often identified with the continuity equation
 \ben\bea{l}
 \nabla \cdot {\bf v}=0.
 \label{div-v}
 \eea\een
This assumption is plausible and indeed consistent with the fluid incompressibility (\ref{div-v}) only if the two fluid components in the mixture are either completely separated by  phase boundaries when their densities are not equal or possibly mixed when the densities are identical. Otherwise, there is a potential inconsistency with the conservation of mass as well as conservation of linear momentum. This
inconsistency has been identified in \cite{LoTr98}, but ignored by many practitioners using phase field modeling technologies for hydrodynamical systems. We note that this inconsistency occurs only in the mixing region of the two incompressible fluids, where the incompressibility condition (\ref{div-v}) is no longer valid, indicating the mixture is no longer incompressible despite that each fluid component participating in mixing is. This type of fluids is  referred to  as quasi-incompressible in \cite{LoTr98}. A systematic fix to this problem for mixtures of incompressible viscous fluids was given by two of the authors in \cite{Li&W2013}, where the divergence free condition is modified to accommodate the quasi-incompressibility.

In modeling of ionic fluids, one recognizes that the size of ions matters in most ionic solutions, in particular in the ionic solutions in which life occurs, in the ocean, and of course in the very crowded conditions found in and near electrodes in batteries and electrochemical cells, in and around enzymes, ionic channels, transporters, and nucleic acids, both DNA and RNA \cite{Zheng&T2015}. Ionic solutions are hardly ever ideal: ionic size is almost always important. In multispecies ionic fluids above a certain concentration or under certain length scales, the size of the ions matters so that the same inconsistency issue in the models for ionic solutions arises again.
That is  one can not simply use the solenoidal condition in the velocity field as a proxy for the material incompressibility. A theory for multispecies ions of incompressible fluid flows that respects the material's mass conservation and momentum conservation needs to be developed.

This paper aims exactly at developing such a theory for a mixture of ionic fluid flows of multiple ionic species, in which the ionic densities are unmatched and different from that of the solvent, and their size effects are non-negligible. We require the theory to be dissipative while conserving mass and momentum. One targeted application of this theory is in  ion channel modeling \cite{Eisenberg1996,Eisenberg1998,Hyon&E&L2011}. Ion channels provide enough data to distinguish between theories because measurements are available over a wide range of conditions \cite{Burger2,Burger1}. Hundreds of channel types are studied every day because of their biological and clinical significance \cite{Zheng&T2015}. Concentrations and electrical potentials are controlled in experiments and these provide sets of values for boundary conditions of mathematical models. Fitting the entire set with one set of structural parameters allows robust solutions of the inverse problem \cite{Burger2,Burger1} and thus allows models to be distinguished. Other applications of the model include electrolyte fluids, biological fluids with charged bio-species etc.  This theory will be consistent with the mass and momentum conservation and demonstrates energy dissipation. In principle, a variety of transport equations can be developed for the ionic species should one knows the system's energy dissipation rate. In this paper, we propose two types of transport equations based on a generalized Onsager principle \cite{Yang&F&W2014}. These two choices yield two types of species transport equations and corresponding energy dissipation rates. Their relations with respect to the existing electrolyte fluid models will be discussed in the text in details.

The derivation follows the generalized Onsager principle approach \cite{Li&W2013,Yang&F&W2014}, leading to two types of transport equations for each ionic species in the form of Cahn-Hilliad and Allen-Cahn type equations, respectively. Apparently, these correspond to two distinct energy dissipation rates. Their applicability to real material systems can only be confirmed if one could measure the systems' energy dissipation rates. However, such measurements have not yet been made, as far as we know. So in most cases, people adopt  one particular formulation over the others simply based on the leap of faith.


 For the new model, together with its limits in the extended Poisson-Nernst-Planck (EPNP) and the classical PNP with the size effect (CPNP), we will study their linearized stability on constant steady states. Instability of the PNP class of models is of direct biological interest. Actual biological channels invariably produce unstable currents \cite{Neher1997} that switch 'instantaneously' between open and closed levels in a random telegraph process called single channel gating \cite{Fitzhugh1983}. Instability in the models of this paper may turn into gating when the models are extended to include noise sources and are focused on the behavior of just one channel protein. However, we will not pursue the complicated issue in this paper; instead, we will focus on introducing the modeling framework and presenting a set of thermodynamically and hydrodynamically consistent theories, and discuss their predictions in a simple 1-D case to highlight the departure of several previously used PNP type models from the new model.

The paper is organized as follows. First we present the mathematical formulation of hydrodynamic phase field theories for multispecies ionic fluid flows and various plausible formulations of the transport equations giving rise to the total energy dissipation. Then, we examine the theory in 1D geometry to compare the theory with some existing   PNP models with and without the size effect \cite{Eisenberg1996,Eisenberg1998,Hyon&E&L2011}. Finally, we provide a concluding remark.

\section{Quasi-incompressible hydrodynamic models for ionic fluids}

\noindent \indent We develop hydrodynamic models for a viscous, multispecies ionic fluid in an isothermal condition, in which mass,  momentum conservation and the total free energy dissipation are preserved. The governing system of equations in the model includes the transport equations for all the ions, the Poisson equation for the electric potential,  and the conservation equation for mass and linear momentum of the fluid, respectively.

\subsection{Mass and momentum conservation equations}

\noindent \indent  We first present the mass and momentum conservation equation. We consider the transport of viscous, ionic fluids made up of $N$ different ionic species, each of which consists of a type of ionic particles of the identical size. Here, we tacitly assume the viscous solvent particle is a type of ions with a zero charge \cite{Lee&etal2013,Tang&etal2001,Lamperski&Z2007,Boda&etal2001}. We denote the number density for each type of ions by $n_i, i=1, \cdots, N$. The electric potential generated by these ionic particles is denoted by $\Phi$. We denote the volume of each individual ionic particle by $v_i$ and the mass by $m_i$ for $i=1,\cdots, N$, respectively. Then, there is a constraint $\sum_{i=1}^N n_iv_i = 1$, which states that the excluded volume of the ions is a constant before and after the mixing.
We identify $i=\alpha$ as the solvent component which is neutral.
The total density of the mixture is defined by
\ben\bea{l}
\rho=\sum_{i=1}^N m_i n_i.
\eea\een
We denote the intrinsic density of the $i$th species by $\rho_i = {m_i}/{v_i}$, which is a constant.
Then, it follows that
\ben
\rho=\sum_{i=1}^N \rho_in_iv_i=\sum_{i=1}^N \phi_i \rho_i,
\een
where $\phi_i=n_iv_i$ is the volume fraction of the ith ion in the mixture.
We introduce the mass averaged velocity ${\bf v}$. Then, the total mass and the linear momentum conservation yield
\ben
\bea{l}
\frac{\partial \rho}{\partial t}+\nabla \cdot (\rho {\bf v}) = 0,\\
\\
\rho (\frac{\partial \bf v}{\partial t} + {\bf v}\cdot\nabla{\bf v}) = \nabla\cdot \tau+{\bf F}^{(e)},
\eea
\een
where $\tau=-p_0{\bf I}+\tau_v$ is the total stress tensor, $p_0$ is the hydrostatic pressure, $\tau_v$ is the extra stress tensor and ${\bf F}^{(e)}$ is the interfacial force that yields the Ericksen stress for the mixture fluid system.
We next turn to the derivation of the transport equations for the ions.

\subsection{Transport equations for the ions}

\noindent \indent The  free energy of the system is prescribed as $F=\int_{\Omega} f[n_1, \cdots, n_N] d{\bf x}$,
where $\Omega$ is the material volume, and the density of the free energy functional is defined by \cite{Rosenfeld&etal1997,Rosenfeld1998}
\ben\bea{l}
f[n_1, \cdots, n_N]=k_BT \sum_{i=1}^N \frac{ n_i}{N_i} (\ln n_i -1)+ \rho^e(\frac{1}{2}\Phi_n+\Phi_e)+\int K({\bf x}-{\bf y}) G(\{n_i\}_{i=1}^N({\bf x}), \{ n_i\}_{i=1}^N({\bf y}))d {\bf y},
\eea\een
where $k_B$ is the Boltzmann constant, $T$ is the absolute temperature, $N_i$ is a generalized polymerization index for the $i$th ionic particle ($N_{\alpha}=1$), $\rho^e = e_0+\sum_{i=1}^N z_i e n_i $ is the total charge density, $z_i$ is the valence for type $i$ ion and also denotes its sign (for solvent, we note that $z_{\alpha}=0$), $e$ is the unit charge, $e_0$ is the permanent charge density in the system, $\Phi_n$ is the electric potential generated by the total charge, $\Phi_e$ is a given external electric potential which is independent of the total charge and the total electric potential is $\Phi = \Phi_n+\Phi_e$. The first group in the sum represents the entropic contribution of the ionic particles to the free energy, the second part gives the electrical energy density of the system, and the third part gives the interaction of the excluded volume effect and the long-range interaction among the ions of finite sizes.

The electrical energy density in the given external electric field is $\rho^e\Phi_e$ and in the electric field generated by the charges is $\frac{1}{2}\rho^e\Phi_n$.
The equations for the electric potentials $\Phi_n$ and $\Phi_e$ are
\ben\bea{l}
\left\{
\bea{l}
\nabla\cdot(\varepsilon\nabla\Phi_n) =  -(e_0 +\sum_i z_i e n_i),\\
\Phi_n|_{\partial\Omega} = 0,
\eea
\right.
\mbox{and}\quad
\left\{
\bea{l}
\nabla\cdot(\varepsilon\nabla\Phi_e) =  0,\\
\Phi_e|_{\partial\Omega} = \Phi_0(\partial\Omega),
\eea
\right.
\eea\een
where $\varepsilon$ is the dielectric constant, $\Phi_0$ is a given boundary function. Here the boundary condition is Dirichlet BC, it can be changed to other type boundary conditions. The external electric potential $\Phi_e$ is determined by the boundary condition with zero charge source. If $\Phi_0 = 0$, there is no external electric potential. $\Phi_n$ is determined by the charge source with homogenous boundary condition and it can be expressed by using the Green's function $G({\bf x}, {\bf x}')$ as
\ben\bea{l}
\Phi_n({\bf x}) = -\int_{\Omega}(G({\bf x}, {\bf x}')(e_0({\bf x}') +\sum_i z_i e n_i({\bf x}')) ) d{\bf x}'.
\eea\een
Then the variation of the electrical energy $F^e = \int_\Omega \rho^e(\frac{1}{2}\Phi_n+\Phi_e) d{\bf x}$ with the ion density $n_i$ is
\ben\bea{l}
\frac{\delta F^e}{\delta n_i} = z_i e \phi_n + z_i e \phi_e = z_i e \Phi.
\eea\een
The equation for the total electric potential is
\ben\bea{l}
\left\{
\bea{l}
\nabla\cdot(\varepsilon\nabla\Phi) =  -(e_0 +\sum_i z_i e n_i),\\
\Phi|_{\partial\Omega} = \Phi_0(\partial\Omega).
\eea
\right.
\eea\een

The third part of the free energy density can be approximated via expansions in a differential form
\ben\bea{l}
\int K({\bf x}-{\bf y}) G(\{n_i\}_{i=1}^N({\bf x}), \{ n_i\}_{i=1}^N({\bf y}))d {\bf y}\approx g[n_1, \cdots, n_N]=g(\{n_i\}_{i=1}^N, \{ \nabla n_i\}_{i=1}^N).
\eea\een
One specific form of the function $g$ accounting for the size effect of the ions is given by
\ben\bea{l}
g=k_BT[\sum_{i,j=1}^N \frac{\xi_{ij}}{2} n_i n_j+\sum_{i=1}^N \frac{\gamma_{i}}{2}\|\nabla n_i\|^2],
\eea\een
where the coefficient matrix $\xi_{ij}$ is symmetry. The first part in the energy density represents a repulsive interaction due to the finite size effect while the second part is the conformation entropy associated with the heterogeneous distribution of the ions in space. This approximate function represents the lowest order approximation to the interaction potential with the long-range interaction, for which we will adopt in the rest of the paper. The chemical potential for the ith ionic particle is then given by
\ben\bea{l}
\mu_i=\frac{\delta F}{\delta n_i}=k_BT[\frac{1}{N_i}(\ln n_i)+ \sum_{j} {\xi_{ij}}n_j - \gamma_{i}\nabla^2 n_i]+e z_i \Phi.
\eea\een

\noindent \indent Assuming there is no annihilation of charges between positive and negative ionic particles, each species' charge  and the total charge in the system is supposed to be conserved under the flux free boundary condition,
\ben\bea{l}
\int_{\Omega} n_i d{\bf x}=C_i, i=1,\cdots, N,\quad \int_{\Omega} (\sum_{i=1}^N z_i  n_i )d{\bf x}=C=const,\label{Charge-neutral}
\eea\een
where $C_i, i=1,\cdots,N$ and $C$ are constants and $C=0$ is called charge neutral. Indeed, annihilation can occur in biological systems and ordinary bulk ionic solutions when weak acids and bases (like acetic acid, i.e., vinegar, or sodium bicarbonate, i.e., baking soda) are involved as components of the solution or as side chains of the protein that forms the ion channel. Such effects are significant in some cases, but they form a separate field of investigation, in theory, experiment, and indeed in medical practice, where they are particularly important. In this paper, we ignore those effects.

We propose the transport equation for the ith ion as follows
\ben
\frac{\partial n_i}{\partial t}+\nabla \cdot ({\bf v} n_i)= B_i, i=1, \cdots, N,
\een
where $B_i$ is  going to be determined from the total free energy dissipation in the following. We note that there are two constraints of $B_i$ as follows, due to the constraint of $n_i$ and the total mass conservation, respectively.
Using $\sum_{i=1}^N n_iv_i = \sum_{i=1}^N \phi_i=1$, we have
\ben
\partial_t(\sum_{i=1}^Nn_iv_i) + \nabla\cdot(\sum_{i=1}^N{\bf v}n_iv_i) = \sum_{i=1}^N B_iv_i.
\een
It implies that
\ben
\bea{l}
\nabla \cdot {\bf v} = \sum_{i=1}^N B_iv_i = \sum_{i=1}^N B_i\frac{m_i}{\rho_i}.
\eea
\een
This gives us the first constraint on the $B_i's$.

In addition, from the total mass conservation and $\rho=\sum_{i=1}^N m_i n_i$, we obtain
\ben
\sum_{i=1}^N m_i B_i=0. \label{B2}
\een
This yields the second constraint on the $B_i's$. The constraints warrants that the transport equations for each species are not completely independent. We next discuss two distinct ways to derive the transport equations for the ions and solvent following the generalized onsager principle \cite{Yang&F&W2014}.

\subsection{Formulation 1}

\noindent \indent We denote the $\alpha$th component (the solvent component) as the non-vanishing component in the mixture and then it follows from eq, (\ref{B2})
\ben\bea{l}
B_{\alpha}=-\frac{1}{m_{\alpha}}\sum_{i\neq \alpha} m_i B_i.
\label{Ba}
\eea\een
The total free energy $E = \int_{\Omega} (\frac{\rho}{2}\|{\bf v}\|^2)d{\bf x} + F$ of the system consists of  two parts: the kinetic energy and the Helmholtz  free energy $F$.  Now, we compute the total free energy dissipation rate as follows:
\ben
\bea{l}
\frac{d E}{dt}=\frac{d}{dt}\int_{\Omega} [\frac{\rho}{2}\|{\bf v}\|^2+ f]d{\bf x}\\
\\
=-\int_{\Omega}[\nabla {\bf v}:\tau-{\bf v}\cdot {\bf F}^{(e)}-\sum_{i=1}^N \mu_i \frac{\partial n_i}{\partial t}] d{\bf x} + \int_{\partial \Omega}{\bf n}\cdot(\sum_{i=1}^N \frac{\partial f}{\partial \nabla n_i}\frac{\partial n_i}{\partial t} )dS\\
\\
=-\int_{\Omega}[\nabla {\bf v}:\tau-{\bf v}\cdot {\bf F}^{(e)}+\sum_{i=1}^N \mu_i (\nabla \cdot {\bf v}n_i+{\bf v}\cdot \nabla n_i)-\sum_{i=1}^N \mu_i B_i] d{\bf x}\\
\\
= -\int_{\Omega}\{\nabla {\bf v}:\tau_v+\sum_{i=1}^N[(-p) m_i (\frac{1}{\rho_i}-\frac{1}{\rho_{\alpha}}) -\mu_i+\frac{m_i}{m_{\alpha}}\mu_{\alpha}] B_i\} d{\bf x},
\eea
\een
where $\partial \Omega$ is the surface of the material volume ${\Omega}$, ${\bf n}$ is the unit external normal, the elastic force is identified as follows
\ben
{\bf F}^{(e)} = \sum_{i=1}^N\mu_i \nabla n_i,
 \een
 and the total pressure is given by
 \ben
 p = p_0 - \sum_{i=1}^N \mu_i n_i.
  \een
  In the last step, constraint eq. (\ref{Ba}) is used. We also set the boundary condition
\ben
{\bf n}\cdot \frac{\partial f}{\partial \nabla n_i} = 0,\label{BC}
\een
so that the surface integration is zero, i.e., $\sum_{i=1}^N\int_{\partial \Omega} {\bf n}\cdot \frac{\partial f}{\partial \nabla n_i} \frac{\partial n_i}{\partial t}ds=0$.

Next, we identify two forms of $B_i$ following the generalized Onsager principle to warrant energy dissipation of the system \cite{Yang&F&W2014}. They are associated with two famous transport equations: the Cahn-Hilliard  and the Allen-Cahn equation, respectively.

\subsubsection{\bf Cahn-Hilliard dynamics}

\noindent \indent In the first case,  we choose $B_i$ as follows
\ben\bea{l}
B_i = -\sum_{k=1}^N \nabla \cdot \lambda_{ik}\nabla  [(-p) m_k (\frac{1}{\rho_k}-\frac{1}{\rho_{\alpha}}) -\mu_k+\frac{m_k}{m_{\alpha}}\mu_{\alpha}], \quad \hbox{for} \quad i\neq\alpha,
\eea\een
where the mobility coefficient matrix $(\lambda_{ij}, i,j\neq\alpha)$ is symmetric and nonnegative definite. Then, using integration by parts, the energy dissipation rate is given by
\ben
\bea{l}
\frac{d E}{dt}=-\int_{\Omega}\{ \nabla {\bf v}:\tau_v+\sum_{i,k=1}^N\nabla  [(-p) m_i (\frac{1}{\rho_i}-\frac{1}{\rho_{\alpha}}) -\mu_i+\frac{m_i}{m_{\alpha}}m_{\alpha}]
\cdot \lambda_{ik}\\\\
\nabla [(-p) m_k (\frac{1}{\rho_k}-\frac{1}{\rho_{\alpha}}) -\mu_k+\frac{m_k}{m_{\alpha}}\mu_{\alpha}] \} d{\bf x} + \hbox{surface term}
\leq 0
\eea
\een
provided ${\nabla {\bf v}}:\tau_v\geq 0$ and the surface term is zero. For viscous fluids, the viscous stress tensor is given by
\ben\bea{l}
\tau_v = 2\eta[{\bf D} - \frac{1}{3}tr({\bf D}){\bf I}] +\nu tr({\bf D}){\bf I},
\eea\een
where ${\bf D} = \frac{1}{2}(\nabla {\bf v}+\nabla {\bf v}^T)$ is the strain rate tensor, ${\bf I}$ is the identity tensor, $\eta$ is the shear viscosity and $\nu$ is the bulk viscosity.
Then ${\nabla {\bf v}}:\tau_v = 2\eta{\bf D}:{\bf D} + (\nu - \frac{2}{3}\eta)(tr({\bf D}))^2 \geq 0$ is satisfied so long as $\eta> 0$ and $\nu - \frac{2}{3}\eta > 0$. The zero surface term is warranted by the following no-flux boundary condition:
\ben\bea{l}
{\bf n}\cdot\{ \sum_{k=1}^N \lambda_{ik} \nabla [(-p) m_k (\frac{1}{\rho_k}-\frac{1}{\rho_{\alpha}}) -\mu_k+\frac{m_k}{m_{\alpha}}\mu_{\alpha}]\}=0.
\label{nof}
\eea\een

We summarize the governing system of equations in this model in the following:
\ben
\bea{l}
\frac{\partial n_i}{\partial t}+\nabla \cdot ({\bf v} n_i) = - \sum_{k=1}^N \nabla \cdot \lambda_{ik}\nabla [(-p) m_k (\frac{1}{\rho_k}-\frac{1}{\rho_{\alpha}}) -\mu_k+\frac{m_k}{m_{\alpha}}\mu_{\alpha}],  \quad \hbox{for} \quad i\neq\alpha,\\
\\
\nabla \cdot {\bf v} = -  \sum_{i,k=1}^N m_i(\frac{1}{\rho_i}-\frac{1}{\rho_{\alpha}}) \nabla \cdot \lambda_{ik}\nabla [(-p) m_k (\frac{1}{\rho_k}-\frac{1}{\rho_{\alpha}}) -\mu_k+\frac{m_k}{m_{\alpha}}\mu_{\alpha}],\\
\\
\rho \frac{d{\bf v}}{dt}=\nabla \cdot [-(p+\sum_{i=1}^N \mu_i n_i){\bf I}+\tau_v]+\sum_{i=1}^N\mu_i \nabla n_i  = \nabla \cdot (-p{\bf I}+\tau_v)-\sum_{i=1}^N n_i \nabla \mu_i,
\eea
\een
and the equation for the electric potential is
\ben\bea{l}
\nabla\cdot(\varepsilon\nabla\Phi) =  -(e_0 +\sum_i z_i e n_i).
\eea\een
where $\varepsilon$ is the dielectric constant. This model is not incompressible since $\nabla \cdot {\bf v}\neq 0$ when densities are not identical. It is known as the quasi-incompressible model \cite{Li&W2014}. This model is different from the previous models for ionic fluids.

We remark that the previous models for ionic fluids assume the incompressible condition
$
\nabla \cdot {\bf v}=0.
$
This is valid only when $\rho_i = \rho_j, i,j=1,\cdots,N$.
In this case, we end up with a self-consistent model as follows:
\ben
\bea{l}
\frac{\partial n_i}{\partial t}+\nabla \cdot ({\bf v} n_i)=\sum_{k=1}^N \nabla \cdot \lambda_{ik}\nabla [\mu_k-\mu_{\alpha}], \quad \hbox{for} \quad i\neq\alpha,\\
\\
\nabla \cdot {\bf v}=0,\\
\\
\rho \frac{d{\bf v}}{dt}=\nabla \cdot (-p{\bf I}+\tau_v)-\sum_{i=1}^N n_i \nabla \mu_i,\\
\\
\nabla\cdot(\varepsilon\nabla\Phi) =  -(e_0 +\sum_i z_i e n_i).
\eea
\een
In this model, the energy dissipation rate is given by
\ben
\frac{d E}{dt}=-\int_{\Omega}\{\nabla {\bf v}:\tau_v+\sum_{i,k\neq\alpha}\nabla [ \mu_i-\mu_{\alpha}] \cdot
\lambda_{ik} \nabla [ \mu_k - \mu_{\alpha}]\}\leq 0.
\een

For the above two model equation systems,   the following boundary conditions are used:
\ben\bea{l}
{\bf n}\cdot \frac{\partial f}{\partial \nabla n_i} = 0,\\
\\
{\bf n}\cdot\{\sum_{k=1}^N \lambda_{ik} \nabla [(-p) m_k (\frac{1}{\rho_k}-\frac{1}{\rho_{\alpha}}) -\mu_k+\frac{m_k}{m_{\alpha}}\mu_{\alpha}]\}=0.
\label{BCn}
\eea\een
Together, they warrant that there is no boundary contribution to the energy dissipation and the constraints on the charge conservation in the system imposed by (\ref{Charge-neutral}) are satisfied.
The boundary condition for the electric potential is the Dirichlet boundary condition which is equal to a specified surface potential, and the boundary condition for the velocity field is the no slip boundary condition.
\\

\subsubsection{\bf Allen-Cahn dynamics}

\noindent \indent Alternatively, we choose $B_i$   as follows
\ben\bea{l}
B_i=\sum_{k=1}^N \lambda_{ik}  [(-p) m_k (\frac{1}{\rho_k}-\frac{1}{\rho_{\alpha}}) -\mu_k+\frac{m_k}{m_{\alpha}}\mu_{\alpha}], \quad \hbox{for} \quad i\neq\alpha,
\eea\een
where $\lambda_{ik}$ is the mobility coefficient, we obtain an Allen-Cahn type transport equation for the ith ion
\ben\bea{l}
\frac{\partial n_i}{\partial t}+\nabla \cdot ({\bf v} n_i)=\sum_{k=1}^N \lambda_{ik}  [(-p) m_k (\frac{1}{\rho_k}-\frac{1}{\rho_{\alpha}}) -\mu_k+\frac{m_k}{m_{\alpha}}\mu_{\alpha}], \quad \hbox{for} \quad i\neq\alpha .
\eea\een
The other equations are given by
\ben
\bea{l}
\nabla \cdot {\bf v} = -  \sum_{i,k=1}^N m_i(\frac{1}{\rho_i}-\frac{1}{\rho_{\alpha}})  \lambda_{ik}  [(-p) m_k (\frac{1}{\rho_k}-\frac{1}{\rho_{\alpha}}) -\mu_k+\frac{m_k}{m_{\alpha}}\mu_{\alpha}],\\
\\
\rho \frac{d{\bf v}}{dt}=\nabla \cdot [-(p+\sum_{i=1}^N \mu_i n_i){\bf I}+\tau_v]+\sum_{i=1}^N\mu_i \nabla n_i  = \nabla \cdot (-p{\bf I}+\tau_v)-\sum_{i=1}^N n_i \nabla \mu_i,
\\\\
\nabla\cdot(\varepsilon\nabla\Phi) =  -(e_0 +\sum_i z_i e n_i).
\eea\een
The boundary condition for this equation system is eq. (\ref{BC}). The energy dissipation rate is given by  the following
\ben
\bea{l}
\frac{d E}{dt}=-\int_{\Omega}\{ \nabla {\bf v}:\tau_v + \\
\sum_{i,k=1}^N [(-p) m_i (\frac{1}{\rho_i}-\frac{1}{\rho_{\alpha}}) -\mu_i+\frac{m_i}{m_{\alpha}}m_{\alpha}]
 \lambda_{ik}
 [(-p) m_k (\frac{1}{\rho_k}-\frac{1}{\rho_{\alpha}}) -\mu_k+\frac{m_k}{m_{\alpha}}\mu_{\alpha}] \} d{\bf x}
\leq 0,
\eea
\een
provided $(\lambda_{ij})\geq 0$.

In the Allen-Cahn model, the charge conservation imposed by (\ref{Charge-neutral}) may not be upheld. In order to impose the constraint approximately, we have to augment the free energy by adding a penalizing term
\ben
L_1\sum_{i=1}^N(\int_{\Omega}n_i-C_i)^2+L_2(\int_{\Omega} \sum_{i=1}^N{z_i n_i}d{\bf x} -C)^2,
\een
where $L_{1,2}$ are  large positive numbers. An alternative approach is to enforce the constraints directly by using Lagrange multipliers in the free energy,
\ben
L_1\sum_{i=1}^N(\int_{\Omega}n_i-C_i)+L_2(\int_{\Omega} \sum_{i=1}^N{z_i n_i}d{\bf x} -C),
\een
where $L_{1,2}$ are two Lagrange multipliers.
These are common practices when one uses Allen-Cahn model to study multiphase fluid dynamics. We note that their physical validity is not widely accepted in the research community though.

Note that Allen-Cahn and Cahn-Hilliard equations represent two different types of transport for scalar phase variables in a dissipative system \cite{Lubensky}. Higher order transport equations are also possible, but are rarely used. Thus, we will not pursue them in this study.

\subsection{Formulation 2}

\noindent \indent By using  constraint eq. (\ref{Ba}), we rewrite the energy dissipation rate as follows
\ben
\bea{l}
\frac{dE}{dt} = -\int_{\Omega}\{\nabla {\bf v}:\tau_v+\sum_{i=1}^N[(-p)  \frac{m_i}{\rho_i} -\mu_i] B_i\} d{\bf x}\\
\\
\quad = -\int_{\Omega}\{\nabla {\bf v}:\tau_v+\sum_{i=1}^N[(-p)  \frac{m_i}{\rho_i} -\mu_i-Lm_i] B_i\} d{\bf x},
\eea
\een
where $L$ is a Lagrange multiplier, which is a function of the space and time. If we adopt the Cahn-Hilliard equation for the ionic species, the right hand term $B_i$ is chosen as
\ben\bea{l}
B_i = -\sum_{j=1}^N\nabla \cdot \lambda_{ij}\nabla [(-p)  \frac{m_j}{\rho_j} -\mu_j-Lm_j], i=1, \cdots, N,
\eea\een
where $\lambda_{ij}$ is the mobility coefficient matrix. The constraint $\sum_{i=1}^N m_iB_i=0$ implies
\ben
\sum_{i,j=1}^N\nabla \cdot \lambda_{ij} \nabla [(-p)  \frac{m_j}{\rho_j} -\mu_j-Lm_j]m_i=0.
\een
It yields an elliptic equation for the  Lagrange multiplier $L$:
\ben\bea{l}
\sum_{i,j=1}^N m_im_j\nabla \cdot \lambda_{ij}\nabla  L=\sum_{i,j=1}^N\nabla \cdot \lambda_{ij} \nabla [(-p)  \frac{m_j}{\rho_j} -\mu_j]m_i.
\eea\een
The Lagrange multiplier $L$ is a solution of the elliptic equation. If the coefficient is a positive definite matrix, $L$ is solvable in principle. In a special case where $\lambda_{ij}$ are constants, the Poisson equation can be rewritten into  \ben
\nabla^2 L = [\sum_{i,j=1}^N \lambda_{ij} m_im_j]^{-1}\sum_{i,j=1}^N\nabla \cdot \lambda_{ij} \nabla [(-p)  \frac{m_j}{\rho_j} -\mu_j]m_i.
\een
 Here, we don't need to know the specific solution form for $L$. Then we have
\ben\bea{l}
B_i = -\sum_{k=1}^N\nabla \cdot \lambda_{ik}\{\nabla [(-p)\frac{m_k}{\rho_k} - \mu_k] - \frac{\sum_{i,j=1}^N \lambda_{ij}m_i m_j \nabla [(-p)  \frac{m_k}{\rho_j} - \frac{m_k}{m_j}\mu_j]}{\sum_{i,j=1}^N \lambda_{ij} m_i m_j} \}\\
\\
\quad = -\sum_{k=1}^N\nabla \cdot \lambda_{ik}G_k.
\eea\een
The flux terms $G_k$ are given by
\ben\bea{l}
G_k = (-\nabla p)(\frac{m_k}{\rho_k} - \frac{\sum_{j=1}^N w_j m_k/\rho_j}{\sum_{j=1}^N w_j}) -\nabla\mu_k + \frac{\sum_{j=1}^N w_j m_k\nabla\mu_j/m_j}{\sum_{j=1}^N w_j}, k = 1,2,...,N.
\eea\een
The terms $w_j=\sum_{i=1}^N \lambda_{ij} m_im_j, j = 1,2,...,N$ act as weighting factors. The difference between this model and the   model derived in formulation 1 is that the correction factors are the weighted average terms.

In a dilute solution, the solvent density is much larger than the other components, that is $n_{\alpha} \gg n_j$ for $j \neq \alpha$. If we assume the mobility parameters $\lambda_{ij} \sim \lambda_i n_i\delta_{ij}$, where $\lambda_i$ is a constant, then $w_j = \sum_{i=1}^N \lambda_{ij} m_im_j \sim \lambda_j n_jm_j^2$. Thus $w_{\alpha} \gg w_j$ for $j\neq \alpha$ when $m_j$ and $m_{\alpha}$ are not far apart, and this formulation reduces to the   Cahn-Hilliard model derived in the previous subsection because
\ben\bea{l}
\frac{\sum_{j=1}^N w_j m_k/\rho_j}{\sum_{j=1}^N w_j} \approx \frac{m_k}{\rho_{\alpha}},\quad
\frac{\sum_{j=1}^N w_j m_k\nabla\mu_j/m_j}{\sum_{j=1}^N w_j}\approx \frac{m_k}{m_{\alpha}}\nabla\mu_{\alpha},\quad
G_{\alpha} \approx 0.
\eea\een
For the solvent component, the governing equation of the density $n_{\alpha}$ is
\ben\bea{l}
\frac{\partial n_{\alpha}}{\partial t}+\nabla \cdot ({\bf v} n_{\alpha}) = - \nabla\cdot \lambda_{\alpha\alpha}G_{\alpha} \approx 0.
\eea\een
Then we can drop the equation of the solvent component in our system and instead only consider the ionic components in this formulation.

If we adopt the Allen-Cahn equation, the $B_i$ is chosen as follows
\ben\bea{l}
B_i=\sum_{j} \lambda_{ij} [(-p)  \frac{m_j}{\rho_j} -\mu_j-Lm_j],
\eea\een
where $\lambda_{ij}$ is the mobility coefficients. The constraint  $\sum_{i=1}^N m_iB_i=0$ implies
$
\sum_{i,j=1}^N\lambda_{ij} [(-p)  \frac{m_j}{\rho_j} -\mu_j-Lm_j]m_i=0.
$
Thus, the Lagrange multiplier $L$ can be solved as follows
\ben\bea{l}
L=[\sum_{i,j=1}^N \lambda_{ij}m_im_j]^{-1}\sum_{i,j=1}^N \lambda_{ij} [(-p)  \frac{m_j}{\rho_j} -\mu_j]m_i.
\eea\een
The transport equation for the ith ion is given by
\ben\bea{l}
\frac{\partial n_i}{\partial t}+\nabla \cdot ({\bf v} n_i) = \sum_{k=1}^N \lambda_{ik}[ (- p)(\frac{m_k}{\rho_k} - \frac{\sum_{j=1}^N w_j m_k/\rho_j}{\sum_{j=1}^N w_j}) -\mu_k + \frac{\sum_{j=1}^N w_j m_k\mu_j/m_j}{\sum_{j=1}^N w_j}].
\eea\een
Using the same argument, if we assume the mobility parameters $\lambda_{ij} \sim \lambda_i n_i\delta_{ij}$, then $w_j = \sum_{i=1}^N \lambda_{ij} m_im_j \sim \lambda_j n_jm_j^2$. Thus, $w_{\alpha} \gg w_j$ for $j\neq \alpha$, which implies
\ben\bea{l}
\frac{\sum_{j=1}^N w_j m_k/\rho_j}{\sum_{j=1}^N w_j} \approx \frac{m_k}{\rho_{\alpha}},\quad
\frac{\sum_{j=1}^N w_j m_k\mu_j/m_j}{\sum_{j=1}^N w_j}\approx \frac{m_k}{m_{\alpha}}\mu_{\alpha},
\eea\een
and the governing equation of the solvent density $n_{\alpha}$ is
\ben\bea{l}
\frac{\partial n_{\alpha}}{\partial t}+\nabla \cdot ({\bf v} n_{\alpha}) \approx \lambda_{\alpha\alpha}[-p(\frac{m_{\alpha}}{\rho_{\alpha}}-\frac{m_{\alpha}}{\rho_{\alpha}})-\mu_{\alpha}+\mu_{\alpha}] = 0.
\eea\een
This formulation reduces to the  Allen-Cahn model derived in formulation 1.


If $(\lambda_{ij})$ is a dense matrix, the two formulations are apparently different. However, if $\lambda_{ij}=\lambda \delta_{ij}$, the Cahn-Hilliard equation derived in formulation 2 reduces to
\ben\bea{l}
\frac{\partial n_i}{\partial t}+\nabla \cdot ({\bf v} n_i) = -  \lambda\nabla^2 [-p\frac{m_k}{\rho_k}+\frac{p}{\sum_{i=1}^N   m_i^2}\sum_{i=1}^N  \frac{m_i^2 m_k}{\rho_i}-\mu_k+\frac{1}{\sum_{i=1}^N   m_i^2}\sum_{i=1}^N   m_i m_k\mu_i].
\eea\een
If $m_i=m, i=1, \cdots, N$, it further reduces to
\ben\bea{l}
\frac{\partial n_i}{\partial t}+\nabla \cdot ({\bf v} n_i) =   \lambda\nabla^2 [ \mu_k-\frac{1}{ N }\sum_{i=1}^N  \mu_i].
\eea\een
Likewise, the Allen-Cahn equation reduces to
\ben\bea{l}
\frac{\partial n_i}{\partial t}+\nabla \cdot ({\bf v} n_i) = -  \lambda [ \mu_k-\frac{1}{ N }\sum_{i=1}^N  \mu_i].
\eea\een
Both of these have been used by some researchers in the past to describe multiphase materials \cite{LoTr98}.

Apparently, formulation 2 is different from formulation 1 and it seems to be a more general way of deriving the transport equations for the ionic species. However, if we choose $L$ such that
\ben
-p\frac{m_{\alpha}}{\rho_{\alpha}}-\mu_{\alpha}-L m_{\alpha}=0
\een
and redefine
\ben
B_{\alpha}=-\frac{1}{m_{\alpha}}\sum_{i\neq \alpha}^N B_im_i,
\een
we recover the model derived using formulation 1. This means that the transport equation for $n_{\alpha}$ defined in reformulation 2 must be modified in order to recover the transport equation in formulation 1. However, this modification has no impact whatsoever on the energy dissipation rate.

Another remark that we would like to make on these models is that each model yields an energy dissipation of its own. The choice of the model should therefore be made based on which energy dissipation rate best fits the real system to be modeled.

\subsection{Model reformulation and reduction to existing models for multispecies ionic fluids}

\noindent \indent The above models are formulated using number densities of the components in the fluid mixture.  We can reformulate the model using the volume fraction $\phi_i$ or the mass fraction $c_i$ since they are functions of the number density functions,
$
\phi_i = n_i \nu_i, \  c_i = \frac{m_i n_i}{\rho},\  i=1,\cdots, N,
$
where $\nu_i$ and $m_i$ are constants, denoting the volume and the mass of each individual ionic particle, respectively.

If $\rho_i=\rho_0, i=1,\cdots,N$, $\nabla \cdot {\bf v}=0$ and,  in addition, we remove the entropic contribution of the solvent to the fluid mixture, i.e., we drop   $n_{\alpha} (\ln n_{\alpha}-n_{\alpha})$, where $\alpha$ corresponds to the solvent component, from the free energy, the model reduces to the existing PNP model with the finite size effect \cite{Hyon&E&L2011,Hyon&E&L2014,Lin&E2015}. So, all the previous ionic fluid models can be regarded as the model applied to the case where all ions are of the same mass density and the solvent effect to the free energy is neglected.

Next, we compare the new model with some of its limits and some existing models.

\section{Binary ionic fluid model}

\noindent \indent We consider a mixture of two distinctive ionic components ($N=3$), where  $\alpha=3$ corresponds to the solvent component, known as the binary ionic fluid model. The other two components in the fluid mixture are  cations (positive ions) and anions (negative ions). We adopt the Cahn-Hilliard dynamics for the transport of ions. The governing system of equations is given by
\ben
\bea{l}
\frac{\partial n_i}{\partial t}+\nabla \cdot ({\bf v} n_i) = - \nabla \cdot \lambda_{i}n_i\nabla [(-p) m_i (\frac{1}{\rho_i}-\frac{1}{\rho_{3}}) -\mu_i+\frac{m_i}{m_{3}}\mu_{3}],  \quad  i=1,2,\\
\\
\nabla \cdot {\bf v} = -  \sum_{i=1}^2 m_i(\frac{1}{\rho_i}-\frac{1}{\rho_{3}}) \nabla \cdot \lambda_{i}n_i\nabla [(-p) m_i (\frac{1}{\rho_i}-\frac{1}{\rho_{3}}) -\mu_i+\frac{m_i}{m_{3}}\mu_{3}],\\
\\
\rho \frac{d{\bf v}}{dt} =
\nabla \cdot (-p{\bf I}+\tau_v)-\sum_{i=1}^3 n_i \nabla \mu_i,\\
\\
\nabla\cdot(\varepsilon\nabla\Phi) =  -(e_0 +\sum_i z_i e n_i),
\eea
\een
Here, we assume the mobility matrix is $\lambda_{ij}=\lambda_i n_i \delta_{ij}$, the mobility of each ion is only dependent on its own number density.
The spatial gradients of the chemical potentials are given by
\ben\bea{l}
\nabla \mu_1 = \frac{k_B T}{N_1 n_1}\nabla n_1 + e z_1 \nabla\Phi + k_BT[\xi_{12}\nabla n_2 + \xi_{11} \nabla n_1 -\gamma_1 \nabla\nabla^2 n_1],\\
\\
\nabla \mu_2 = \frac{k_B T}{N_2 n_2}\nabla n_2 + e z_2 \nabla\Phi + k_BT[\xi_{12}\nabla n_1 + \xi_{22} \nabla n_2 -\gamma_2 \nabla\nabla^2 n_2],\\
\\
\nabla\mu_3 = k_B T \frac{-v_1 \nabla n_1 - v_2 \nabla n_2}{1-v_2 n_2 -v_1 n_1}.
\eea\een
Where we assume that $\xi_{3i} = \xi_{j3} = \gamma_3 = 0$, i.e., the interaction between the ions is dominant. The entropic contribution only shows up in the  chemical potential of solvent ($\mu_3$).

\subsection{Nondimensionalization}

\noindent \indent We use a characteristic time scale $t_0$, length scale $l_0$, and mass density scale $\rho_0=\rho_3$, and the characteristic number density $n_0$  to non-dimensionalize the physical variables. The mass density scale is chosen as the mass density of water here. Then, we denote the corresponding volume scale as $v_0 = l_0^3$, mass scale $m_0 = \rho_3 v_0$. The dimensionless variables are defined as follows:
\ben
\tilde{n}_i = \frac{n_i}{n_0}, \   \tilde{t} = \frac{t}{t_0},\  \tilde{x} = \frac{x}{l_0},\  \tilde{\bf v} = \frac{t_0}{l_0}{\bf v},\  \tilde{p} = \frac{ t_0^2}{\rho_0l_0^2}p, \  \tilde{\mu}_i = \frac{ t_0^2}{m_0l_0^2}\mu_i.
\een
Then, the dimensionless parameters are given by
\ben\bea{l}
r^{v}_i = \frac{v_i}{v_3},\ r^{m}_i = \frac{m_i}{m_3},i=0,1,2, \quad \tilde{\rho} = \frac{\rho}{\rho_0}, \quad \tilde{\rho}_i = \frac{\rho_i}{\rho_0}, \quad \tilde{\lambda}_i = \frac{m_0}{t_0}\lambda_i,\quad \tilde{\eta} = \frac{t_0}{\rho_0 l_0^2}\eta, \quad \tilde{\nu} = \frac{t_0}{\rho_0 l_0^2}\nu,\\
\\
\tilde{k}_B = \frac{Tt_0^2}{m_0l_0^2}k_B, \quad
\tilde{\xi}_{ij} = n_0\xi_{ij}, \quad \tilde{\gamma}_{i} = \frac{n_0}{l_0^2}\gamma_{i}, \quad \tilde{\Phi} = \frac{t_0^2 e}{m_0l_0^2}\Phi,\quad \tilde{e}_0 = \frac{e_0}{e n_0},\quad \tilde{z}_i = {z_i}, \quad \tilde{\varepsilon} = \frac{m_0}{e^2t_0^2n_0}\varepsilon.
\eea\een
We set $\tilde k_B=1$ to obtain $t_0=\sqrt{\frac{m_0 l_0^2}{k_BT}}$ and also set $n_0 v_0=1$ to obtain $n_0=\frac{1}{v_0}$. It's easy to find that $r_i^m = \tilde{\rho}_i r_i^v$ for $i=1,2$ and $r_0^m = r_0^v$.
For simplicity, we drop the $\ \tilde{}\ $ on the dimensionless variables and the parameters. The system of governing equations for the  binary ionic fluid model in these dimensionless variables are given by
\ben\bea{l}
\frac{\partial n_i}{\partial t}+\nabla \cdot ({\bf v} n_i) = - \nabla \cdot \lambda_{i}n_i\nabla [ - R_i p -\mu_i + {r^m_i}\mu_{3}],  \quad  i=1,2,\\
\\
\nabla \cdot {\bf v} = -  \sum_{i=1}^2 R_i \nabla \cdot \lambda_{i}n_i\nabla [ - R_i p -\mu_i + {r^m_i}\mu_{3}]
 = \sum_{i=1}^2 R_i [\frac{\partial n_i}{\partial t}+\nabla \cdot ({\bf v} n_i)],\\
\\
\rho \frac{d{\bf v}}{dt}  = \nabla \cdot (-p{\bf I}+\tau_v)- \sum_{i=1}^3 n_i \nabla \mu_i,\\
\\
 \nabla\cdot(\varepsilon\nabla\Phi) = -(e_0 +\sum_{i=1}^2 z_i n_i),
\eea\een
where the parameters $R_i = (\frac{r^m_i}{r_0^m})(\frac{1}{\rho_i}-1)$ for $i=1,2$, the total mass density $\rho = 1 -R_1n_1 -R_2n_2$, the solvent's number density $n_3 = {r_0^v}- r^v_1 n_1-r^v_2 n_2$. The spatial gradients of the chemical potentials are
\ben\bea{l}
\nabla \mu_1 = \frac{1}{N_1 n_1}\nabla n_1 + z_1 \nabla \Phi + \xi_{12}\nabla n_2 + \xi_{11} \nabla n_1 -\gamma_1\nabla\nabla^2n_1,\\
\\
\nabla \mu_2 = \frac{1}{N_2 n_2}\nabla n_2 + z_2 \nabla\Phi + \xi_{12}\nabla n_1 + \xi_{22} \nabla n_2 -\gamma_2\nabla\nabla^2n_2,\\
\\
\nabla\mu_3 =  \frac{-r^v_1 \nabla n_1 - r^v_2 \nabla n_2}{n_3}.
\eea\een
In the following, we refer to the  model as the full model, where the word "full" means that the model respects all conservation laws and accounts for the finite size effect and the solvent entropy.

\subsection{Models at regimes of two distinct length scales}

\noindent \indent We examine the dimensionless full model at two distinct length scales.
If we choose the length scale $l_0 = 10^{-9}m = 1 nm$,  we have the time scale  $t_0 = 1.55\times 10^{-11}$s.
If we choose the length scale $l_0 = 10^{-7}m = 100 nm$,  we have the time scale $t_0 = 1.55\times10^{-6}s $.

We set the first  type ion is the positive ion with valence $z_1 = +1$ and polymerization index $N_1 =1$; the second type ion is the negative ion with valence $z_2 = -1$ and polymerization index $N_2 = 1$. The values of the ratios of the ions' volume, mass and density are tabulated in Table 1. The density ratio of the solvent and two ions is $\rho_3:\rho_1:\rho_2 = 1:0.5:2$, the volume ratio is $v_3:v_1:v_2=1:2:1$. The size differences of the three components are distinct. The parameters $R_1,R_2$ are $O(10^{-2})$ in the smaller length scale $l_0 = 1nm$. The compressibility of the flow ($\nabla\cdot{\bf v}\neq 0$) in the full model can not be neglected.
\begin{table}[!thp]
\centerline{Table 1: The ratios of volume, mass and density}
\vskip 12 pt
\centering
\begin{tabular}{c|c|c|c|c|c|c}
 \cline{1-7} Ratios & $\rho_1$ & $\rho_2$ & $r_1^v$ & $r_2^v$ & $r_1^m$ & $r_2^m$ \\
 \cline{1-7}
           Values & 0.5&  2 & 2 & 1 & 1 & 2 \\
 \cline{1-7} Ratios &\multicolumn{2}{|c|}{$r_0^v = r_0^m$}  &\multicolumn{2}{|c|}{$R_1$} & \multicolumn{2}{|c}{$R_2$} \\
  \cline{1-7}
 Values $(l_0 = 1 nm)$ & \multicolumn{2}{|c|}{40} & \multicolumn{2}{|c|}{0.025} & \multicolumn{2}{|c}{$-0.025$} \\
  Values $(l_0 = 100nm)$ & \multicolumn{2}{|c|}{$4\times10^7$}& \multicolumn{2}{|c|}{$  2.5\times10^{-8}$} & \multicolumn{2}{|c}{$-2.5\times10^{-8}$} \\
 \cline{1-7}
\end{tabular}
\end{table}

If the densities of the three components are the same, i.e. $\rho_3:\rho_1:\rho_2 = 1:1:1$, we have $R_1 = R_2 = 0$,  the flow becomes incompressible. Furthermore, when the density differences are distinct, but the larger characteristic length scale $l_0 = 100nm$ is used in the dimensionless system, the values of parameters $R_1,R_2$ are very small, as in Table 1. If the corresponding terms of $R_i$ are dropped from the full model, the model reduces to a model that we call the extended PNP model (EPNP), in which the flow is incompressible:
\ben\bea{l}
\frac{\partial n_i}{\partial t}+\nabla \cdot ({\bf v} n_i) = - \nabla \cdot \lambda_{i}n_i\nabla [ -\mu_i + {r^m_i}\mu_{3}],  \quad  i=1,2,\\
\\
\nabla \cdot {\bf v} = 0,\\
\\
\rho \frac{d{\bf v}}{dt}  = \nabla \cdot (-p{\bf I}+\tau_v)- \sum_{i=1}^3 n_i \nabla \mu_i,\\
\\
 \nabla\cdot(\varepsilon\nabla\Phi) = -(e_0 +\sum_{i=1}^2 z_i n_i).
\eea\een
Here the mass density, the stress tensor and the chemical potentials are the same to those in the  Full model.

Furthermore, if we neglect the entropic contribution of the solvent to the fluid mixture, i.e., we drop $n_{\alpha}(ln n_{\alpha} -1 ), \alpha=3,$ from the free energy, we get the classical PNP model with the finite size effect (CPNP) \cite{Eisenberg1996,Eisenberg1998,Eisenberg2000,Eisenberg2001,Eisenberg2002}. This is equivalent to removing the $\mu_3$ terms from the equations of the above EPNP model. To be clear, we note that the commonly used classical PNP model does not include the finite size effect.

When the characteristic length scale used is $l_0=1nm$, the parameters $R_1,R_2$ are not small. So, the full model must be used. The model is indeed different from the limiting PNP models even with the finite size effect. Note that this is the length scale regime that is applicable to the ion channel problem. The other model parameters are list in Table 2.

\begin{table}[!thp]
\centerline{Table 2: Model Parameters}
\vskip 12 pt
\centering
\begin{tabular}{l|l|l|l|l}
 \cline{1-5} Symbol & Parameter & Value (Unit) &  $l_0 = 1nm$ &  $l_0 = 100nm$\\
\cline{1-5}
$\eta$        & Shear viscosity    & $ 1\times10^{-3}$ $kg m^{-1}s^{-1}$    & 15.54 &  155.4\\
$\nu$         & Bulk viscosity     & $ 2.75\times10^{-3}$ $kg m^{-1}s^{-1}$ & 42.74  &  427.4\\
$\varepsilon$ & Dielectric constant  & $7.08\times 10^{-10} F m^{-1}$ & 0.1145 &  11.45\\
$\Phi$        & Electric potential  & $ 1 V$ & 38.65  & 38.65 \\
\cline{1-5}
$\gamma_1$    & High order diffusion  of 1th ion  & $ 1.6606\times10^{-27}$ $ m^5 mol^{-1}$ & $10^{-4}$ & $10^{-14}$\\
$\gamma_2$    & High order diffusion  of 2th ion  & $ 1.6606\times10^{-27}$ $ m^5 mol^{-1}$ & $10^{-4}$ & $10^{-14}$\\
$\lambda_1$   & Mobility of 1th ion  & $ 3.1083\times10^{11}$ $ kg^{-1}s$ & $0.02$ & $0.2$\\
$\lambda_2$   & Mobility of 2th ion  & $ 3.1083\times10^{11}$ $ kg^{-1}s$ & $0.02$ & $0.2$\\
$\xi_{11}$    & Self-interaction of 1th ion  & $ 1.6606\times10^{-5}$ $ m^{3}mol^{-1}$ & 1 & $10^{-6}$\\
$\xi_{22}$    & Self-interaction of 2th ion  & $ 1.6606\times10^{-5}$ $ m^{3}mol^{-1}$ & 1 & $10^{-6}$\\
$\xi_{12}$    & Interaction of the two ions  & $ 1.6606\times10^{-5}$ $ m^{3}mol^{-1}$ & 1 & $10^{-6}$\\
\cline{1-5}
\end{tabular}
\end{table}

\subsection{ Comparison of the full model with the limiting PNP models in 1D space}

\noindent \indent We compare the full model with the EPNP and CPNP models in 1D space, assuming the system is homogeneous in the $(y,z)$ directions and depends only on $x$ and time $t$ (i.e., the variables are functions of $(t,x)$.) The domain for $x$ is assumed finite given by $\Omega = [0,L_x]$. The governing equations of full model in 1D are given explicitly by
\ben\bea{l}
\frac{\partial n_1}{\partial t}+ ({\bf v}_1 n_1)'= -\{ \lambda_{1}n_1 [-R_1(p)' -(\mu_1)'+ {r^m_1}(\mu_{3})']\}',\\
\\
\frac{\partial n_2}{\partial t}+ ({\bf v}_1 n_2)'= -\{ \lambda_{2}n_2 [-R_2(p)' -(\mu_2)'+ {r^m_2}(\mu_{3})']\}',\\
\\
({\bf v}_1)'  
= R_1[\frac{\partial n_1}{\partial t}+ ({\bf v}_1 n_1)'] + R_2[\frac{\partial n_2}{\partial t}+ ({\bf v}_1 n_2)'],\\
\\
\rho(\frac{\partial {\bf v}_1}{\partial t} + {\bf v}_1({\bf v}_1)') = -(p)'+(\frac{4}{3}\eta+\nu)({\bf v}_1)'' - [ n_1(\mu_1)'+n_2(\mu_2)'+n_3(\mu_3)'],\\
\\
\rho(\frac{\partial {\bf v}_2}{\partial t} + {\bf v}_1({\bf v}_2)') = \eta({\bf v}_2)'',\\
\\
\rho(\frac{\partial {\bf v}_3}{\partial t} + {\bf v}_1({\bf v}_3)') = \eta({\bf v}_3)'',\\
\\
(\Phi)'' = -[e_0 +\sum_{i=1}^2 z_i n_i]/\varepsilon,
\label{govern-eq}
\eea\een
where $(\cdot)' = \frac{\partial (\cdot)}{\partial x}, (\cdot)'' = \frac{\partial^2 (\cdot)}{\partial x^2}$, and the gradients of the chemical potentials are
\ben\bea{l}
(\mu_1)' = \frac{1}{N_1 n_1} (n_1)' + z_1 (\Phi)' + \xi_{12}(n_2)' + \xi_{11} (n_1)' - \gamma_1(n_1)''',\\
\\
(\mu_2)' = \frac{1}{N_2 n_2} (n_2)' + z_2 (\Phi)' + \xi_{12}(n_1)' + \xi_{22} (n_2)' - \gamma_2(n_2)''',\\
\\
(\mu_3)' =  \frac{-r^v_1 (n_1)' - r^v_2 (n_2)' }{n_3}.
\label{cheme}
\eea\een
The unknowns are $n_1, n_2, p, {\bf v}_1, {\bf v}_2, {\bf v}_3, \Phi$, which are fully coupled.

The 1D EPNP model is much simpler now. First, from the incompressible condition $({\bf v}_1)' = 0$, we find that ${\bf v}_1 = 0$ due to the fixed boundary condition of ${\bf v}$. Then, the pressure $p$ and ${\bf v}_2, {\bf v}_3$ are determined from the momentum equation. The independent unknowns in the EPNP model are then $n_1, n_2, \Phi$ and the 1D governing equations are given by
\ben\bea{l}
\frac{\partial n_1}{\partial t} = - \{ \lambda_{1}n_1 [-(\mu_1)' + {r^m_1}(\mu_{3})']\}',\\
\\
\frac{\partial n_2}{\partial t} = - \{ \lambda_{2}n_2 [-(\mu_2)' + {r^m_2}(\mu_{3})']\}',\\
\\
(\Phi)'' = -[e_0 +\sum_{i=1}^2 z_i n_i]/\varepsilon,
\label{govern-eq-pnp}
\eea\een
where the gradients of the chemical potentials are given by  (\ref{cheme}).

If we further remove the $\mu_3$ terms from the 1D governing equations of the EPNP model, we get the 1D equations of the classical PNP model with the finite size effect (CPNP). In the following, we will compare these three models explicitly in 1D. First, we examine their linear stability properties.

\subsection{Linear stability of the constant state}

\noindent \indent If we assume $e_0 = 0$ (namely, the system does not have a permanent charge present), there exists a constant solution of the full model, which is a solution of all limiting models, $n_1= n_2 = n^0, \ {\bf v} = {\bf 0}, \ p = 0, \  \Phi = 0$, where $n^0$ is constant such that
\ben\bea{l}
n_3^0 = r_0^v-r_1^v n^0 - r_2^v n^0 > 0.
\label{solv}
\eea\een
This inequality is necessary to ensure that the solvent density is greater than zero.
We perturb this constant solution as follows:
\ben\bea{l}
n_1 = n^0 + \epsilon e^{\alpha t + ikx} n_1^0, \quad n_2 = n^0 + \epsilon e^{\alpha t + ikx} n_2^0,\\
{\bf v}_1 = \epsilon e^{\alpha t + ikx} {\bf v}_1^0, \quad p = \epsilon e^{\alpha t + ikx} p^0, \quad
\Phi = \epsilon e^{\alpha t + ikx} \Phi^0.
\eea\een
Here $\epsilon \ll 1$ is a small parameter, $\alpha$ is the growth rate and $k$ is the wave number.
First, we point out that the velocity components ${\bf v}_2, {\bf v}_3$ are decoupled from the rest of the system in the linearized equations and they do not contribute instability in this problem; so, we only consider the coupled system involving the remaining variables: $p^0, \Phi^0, {\bf v}_1^0, n_1^0, n_2^0$.
The linearized eigenvalue problem for the Full model is given in the Appendix. The asymptotical analysis in the small wave number regime shows that the instability can incur only when  $\xi_{12}$ is negative enough. But, $\xi_{ij}>0$ in the model. So this mode of instability is absent from the full model and its limits. The system is stable for long wave (small wave number) perturbation. It is easy to find that the system is also stable for short wave (large wave number) perturbation. From the numerical studies, we find that the intermediate wave instability appears when $\xi_{12}$ is sufficiently large. The analytical result of the intermediate wave instability is hard to obtain from  the full model, but easy from the EPNP model. We thus focus on the linear stability of the limiting EPNP model in the following.

The linearized eigenvalue problem is given in the Appendix. The instability condition is $A<0$, where
\ben\bea{l}
 A = \lambda_1\lambda_2(n^0)^2 \left\{ak^2 + bk^4 + [\gamma_2(\frac{1}{N_1 n^0} + \xi_{11} +\frac{r_1^vr_1^m}{n_3^0}) + \gamma_1(\frac{1}{N_2 n^0} + \xi_{22} +\frac{r_2^vr_2^m}{n_3^0})]k^6+\gamma_1\gamma_2k^8  \right\}<0.
\label{A}
\eea\een
Here, the parameters $a,b$ are defined by
\benl\bea{l}
a = [\frac{1}{N_1 n^0} + \frac{1}{N_2 n^0} + \xi_{11}+\xi_{22}+ 2\xi_{12} + \frac{1}{n_3^0}(r_1^v+r_2^v)(r_1^m+r_2^m)]\frac{1}{\varepsilon},\\
b = \frac{\gamma_1+\gamma_2}{\varepsilon}
+ \frac{1}{N_1N_2 (n^0)^2} + (\xi_{11}+\frac{r_1^vr_1^m}{n_3^0})\frac{1}{N_2 n^0} + (\xi_{22} + \frac{r_2^vr_2^m}{n_3^0})\frac{1}{N_1 n^0} + \\
\quad \frac{r_1^vr_1^m\xi_{22}+ r_2^vr_2^m\xi_{11}}{n_3^0} + \xi_{11}\xi_{22} - \xi_{12}^2 - \frac{r_1^mr_2^v+r_2^mr_1^v}{n_3^0}\xi_{12}.
\eea\eenl
Because $\lambda_1, \lambda_2, \gamma_1, \gamma_2, \xi_{11}, \xi_{22}$ are all positive, so the coefficients of $k^8, k^6$ are all positive. It implies that $A > 0$ for large wave numbers. This means that the system is stable for short waves. For long waves (small wave numbers), since the parameter $a>0$,
then $A>0$ for small $|k|$. So, the solution is stable.   Analogously, the mode of instability is absent from the CPNP model, which is a limit of the EPNP model, at both long and short waves.

\begin{figure}[!ht]
\centering
\subfigure[Fixed $n^0 = 1$]       {\includegraphics[width=0.45\textwidth]{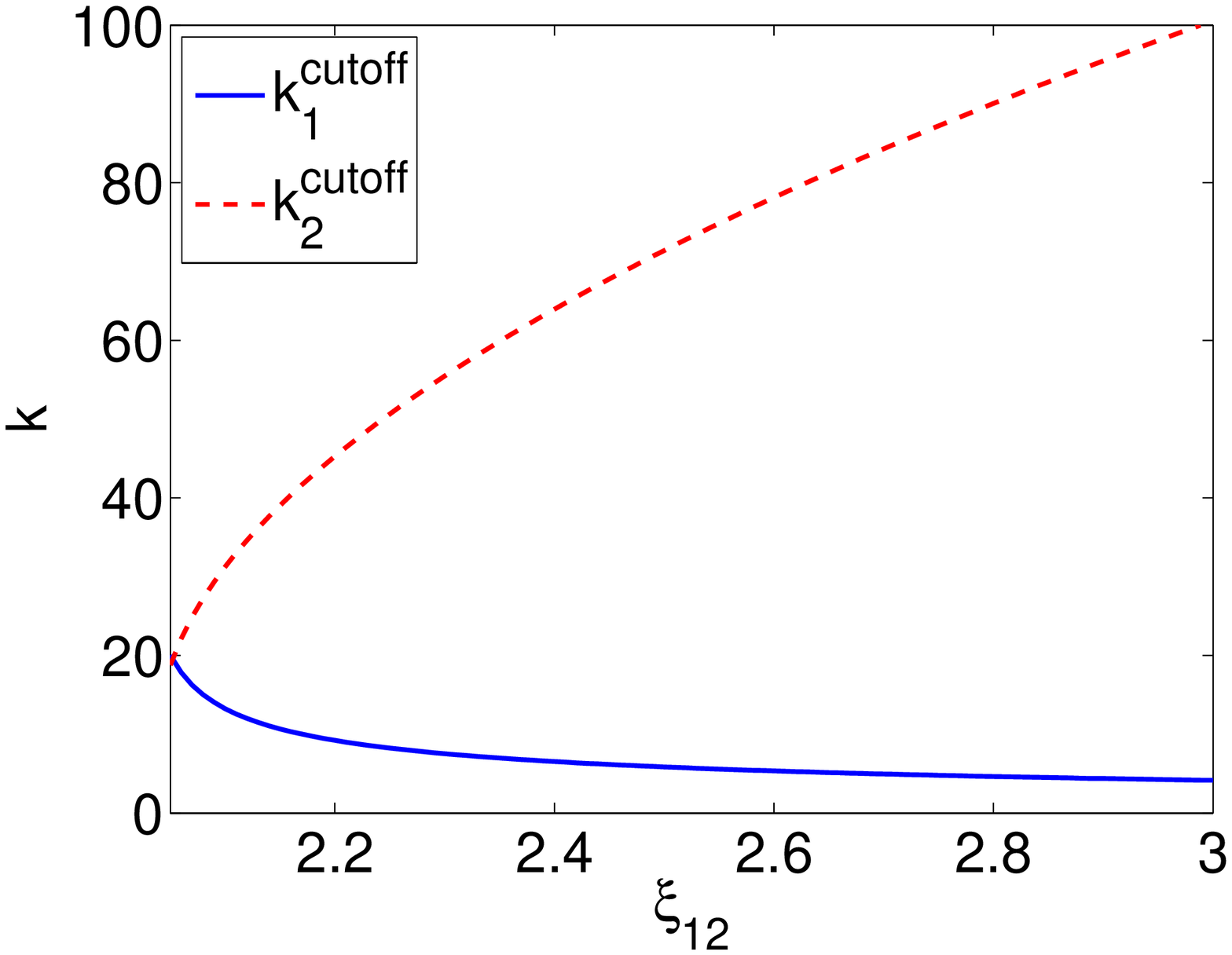}}
\subfigure[Fixed $\xi_{12} = 2.2$]{\includegraphics[width=0.45\textwidth]{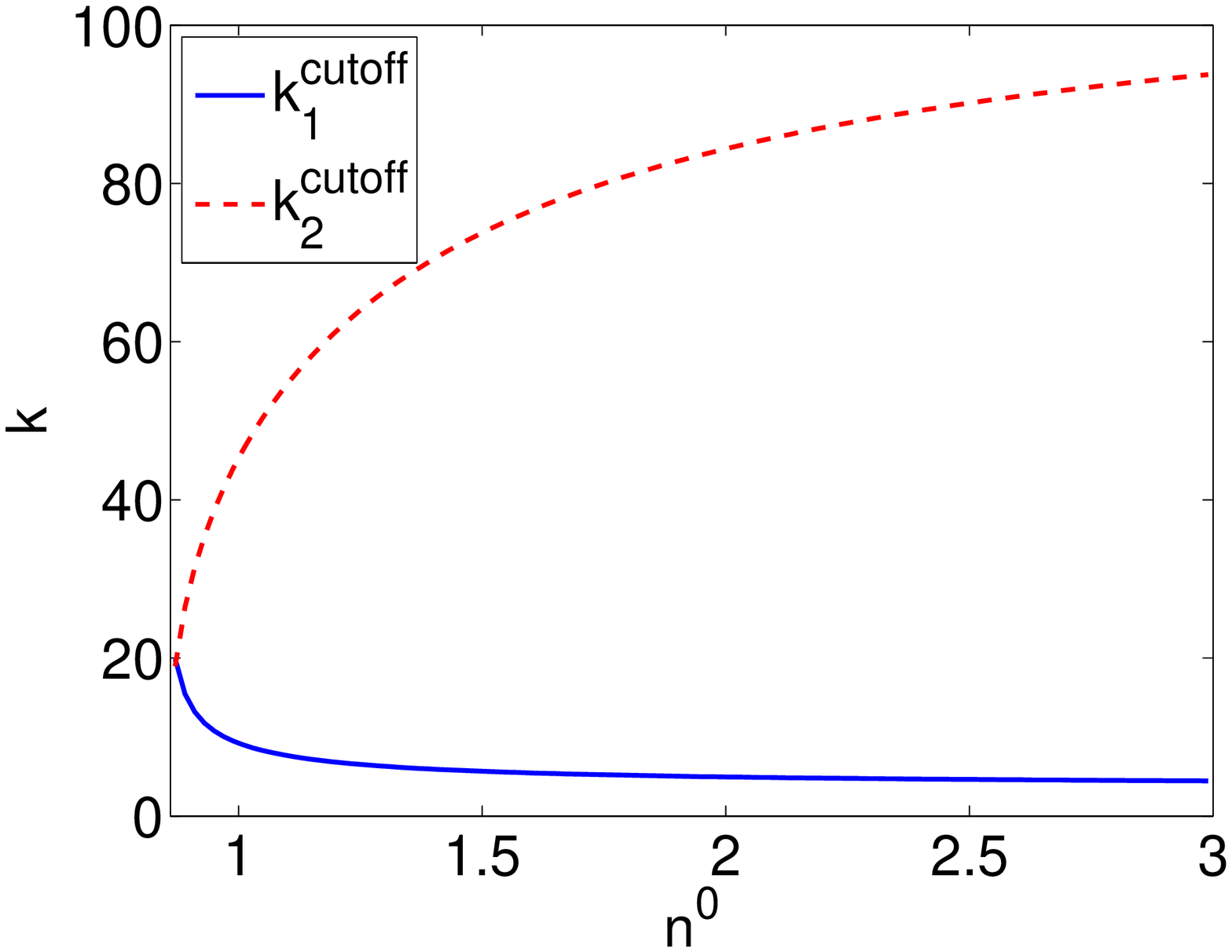}}
\caption{Cutoff wave numbers as functions of $\xi_{12}$ with $n^0=1$ and as functions of $n^0$ with $\xi_{12} = 2.2$. }
\label{cut}
\end{figure}

For intermediate waves, we notice a possible instability if $b$ is negative, i.e.,
\ben\bea{l}
a > 0, \quad b < 0.
\label{inst}
\eea\een
In certain parameter regimes, the growth rate $\alpha_1$ (given in the Appendix) can be negative for a very small $|k|$, becomes positive for some intermediate values of $|k|$, and then turns to negative again at large $|k|$. Assuming $\gamma_1=\gamma_2 = \delta \ll 1$, we obtain the  roots of  $A=0$ asymptotical. Then, we obtain the cutoff wave numbers.
We denote $c = \frac{1}{N_1 n^0} + \xi_{11} +\frac{r_1^vr_1^m}{n_3^0}+ \frac{1}{N_2 n^0} + \xi_{22} +\frac{r_2^vr_2^m}{n_3^0}$ , then we have
\benl\bea{l}
A/(\lambda_1\lambda_2(n^0)^2k^2) = a + bk^2 + c\delta k^4 + \delta^2 k^6.
\eea\eenl
There are two positive roots of $k^2$, corresponding to two cutoff wave numbers $k_{1,2}^{cutoff}$, asymptotically:
\ben\bea{l}
(k_1^{cutoff})^2 = -\frac{a}{b} - \frac{a^2c}{b^3}\delta + O(\delta^2), \quad (k_2^{cutoff})^2 = \frac{x_0}{\delta} + x_1 + O(\delta),
\eea\een
where $x_0 = \frac{-c + \sqrt{c^2-4b}}{2}, x_1 = \frac{-a}{b+2cx_0+3x_0^2}$. We only retain the first two  terms in the asymptotic roots.
The parameter $A$ is negative when the wave number is between the two cutoff wave numbers $0<k_1^{cutoff} < k < k_2^{cutoff}$. The growth rate $\alpha_1$ is positive in this intermediate wave number regime. This instability depends strongly on the interaction parameter $\xi_{12}$, the instability condition is satisfied for a sufficiently large $\xi_{12}$. We fix the parameter $N_1=N_2 = 1, \lambda_1 = \lambda_2 = 0.02, \gamma_1 = \gamma_2 = 10^{-4}, \xi_{11} = \xi_{22} = 1, n^0 = 1$ in the length scale $l_0 = 1nm$ regime, and vary the parameter $\xi_{12}$. When $\xi_{12}>2.06$, $b<0$, the intermediate wave instability incurs. In Figure \ref{cut}(a), we show the curves of the two cutoff wave numbers as a function of $\xi_{12}$. The smaller cutoff wave number $k_1^{cutoff}$ is decreasing and the larger cutoff wave number $k_2^{cutoff}$ is increasing as $\xi_{12}$ increases from $2.046$. The unstable wave number regime $(k_1^{cutoff}, k_2^{cutoff})$ widens as $\xi_{12}$ increases.

When $\delta \rightarrow 0$, we have $k_1^{cutoff} \rightarrow \sqrt{-\frac{a}{b}}$ and $k_2^{cutoff} \rightarrow + \infty$. The unstable wave regime is $k>k_1^{cutoff}$. The system is unstable for large wave numbers (short waves), which is known as the Hadamard instability. Hence, the high order diffusion coefficients $\gamma_1,\gamma_2$ have the effect to suppress the short wave instability.

This intermediate wave instability is also dependent of the constant state $n^0$. When the interaction parameter $\xi_{12} = 2.2$ is fixed, but $n^0$ is varying, we find that $b$ is positive for small $n^0$ and negative for large $n^0$. That means the system is stable for dilute solution but unstable for rich solution.
We also plot the cutoff wave numbers as functions of $n^0$ with fixed $\xi_{12} = 2.2$ in Figure \ref{cut}(b). The instability appears when $n^0>0.87$, and the unstable wave number regime $(k_1^{cutoff}, k_2^{cutoff})$ widens as $n^0$ increases.

\begin{figure}[!ht]
\centering
\subfigure[$l_0 = 1nm.  $]{\includegraphics[width=0.325\textwidth]{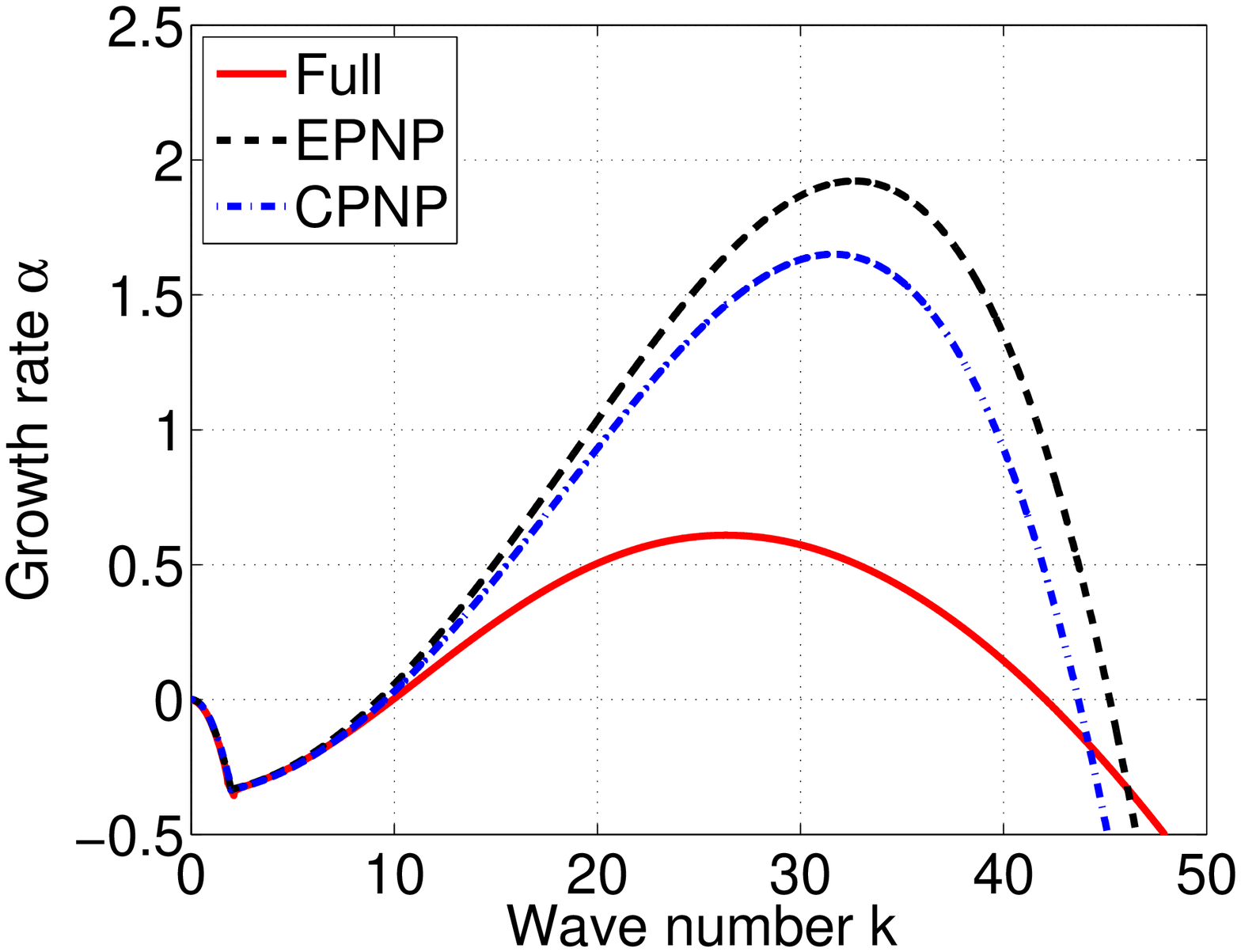}}
\subfigure[$l_0 = 10nm. $]{\includegraphics[width=0.325\textwidth]{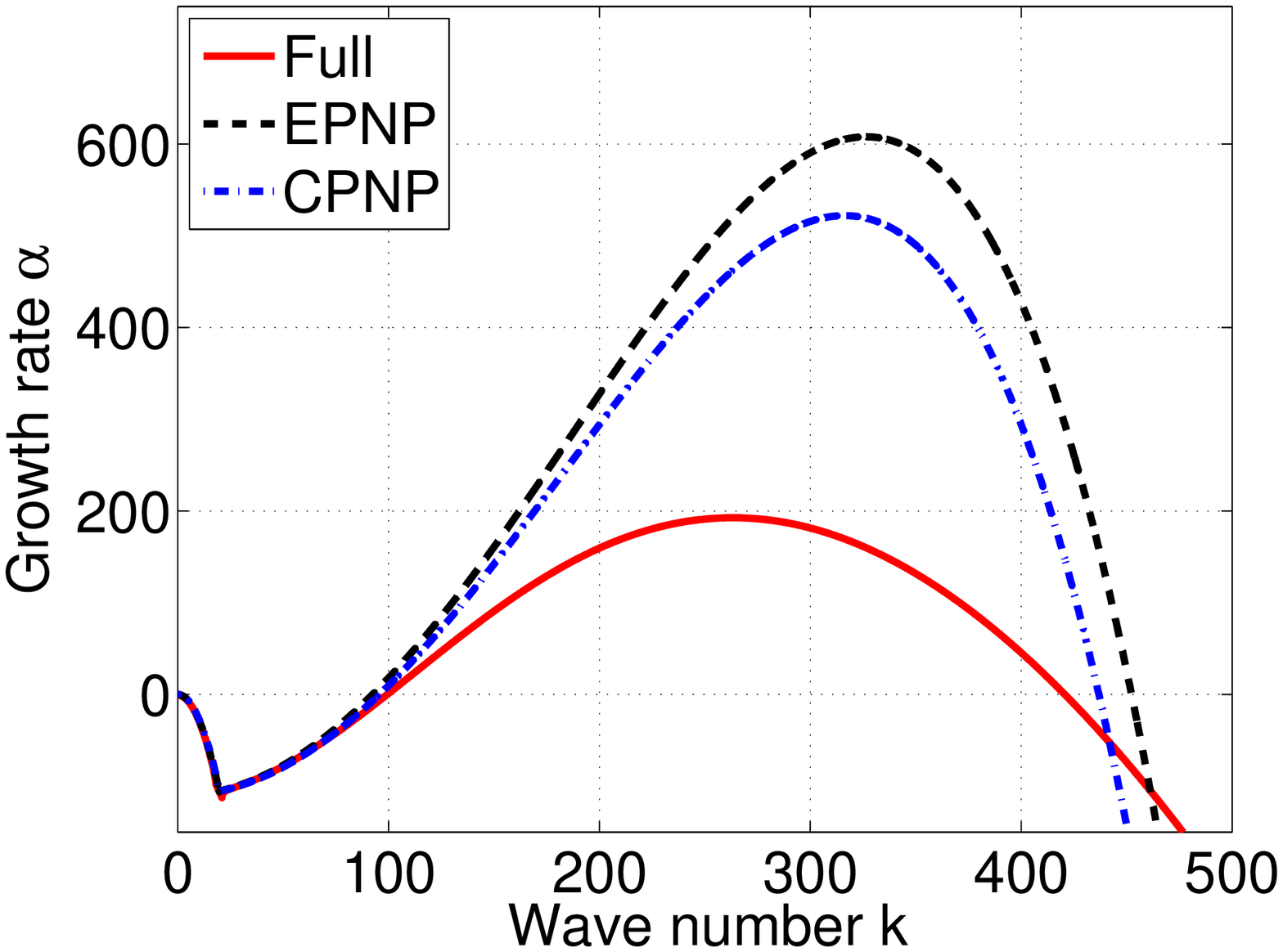}}
\subfigure[$l_0 = 100nm.$]{\includegraphics[width=0.325\textwidth]{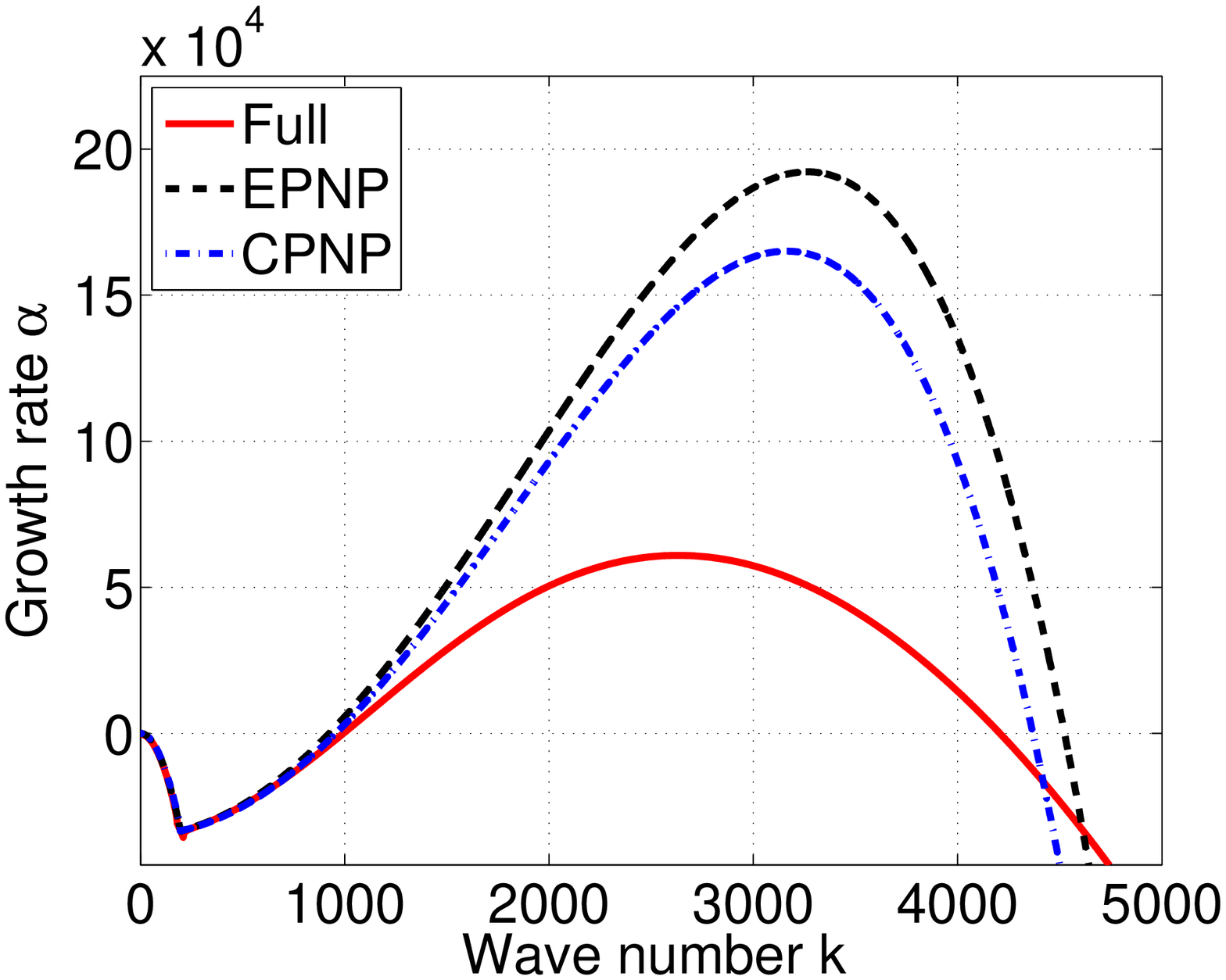}}
\caption{The growth rates of the full model and the two PNP models with length scale $l_0 = 1,10,100nm$, respectively, in the parameter regime of intermediate wave number instability. The values of growth rates increase as the length scale $l_0$ increases. The two cutoff wave numbers of the EPNP model in the length scale $l_0 = 1nm$ regime are $9.24$ and $45.23$, respectively. The full model is more stable than the other two models in this regime. }
\label{Tgrowth4}
\end{figure}

This intermediate wave instability is a feature of these three models. Through a numerical investigation, we confirm that this instability property can occur in all three models. In the following example (Figure \ref{Tgrowth4}), we use parameter values $N_1=N_2 = 1, \lambda_1 = \lambda_2 = 0.02, \gamma_1 = \gamma_2 = 10^{-4}, \xi_{11} = \xi_{22} = 1, \xi_{12} = 2.2, n^0 = 1$ in the length scale $l_0 = 1nm$ regime. The instability condition (\ref{inst}) is satisfied. The two asymptotical cutoff wave numbers of EPNP model are $k_1^{cutoff} = 9.24$ and $k_2^{cutoff} = 45.23$, respectively, when length scale $l_0 = 1nm$. For the three models, the relation between the length scale and the growth rate follows a simple scaling law: we denote the two length scales as $l_0^{(1)}, l_0^{(2)}$, the corresponding growth rates as $\alpha_1^{(1)}, \alpha_1^{(2)}$, and the cutoff wave numbers as $k^{(1)}, k^{(2)}$, respectively. If $\frac{l_0^{(2)}}{l_0^{(1)}} = K$, then the cutoff wave number ratio follows  $\frac{k^{(2)}}{k^{(1)}} = K$ while the growth rate ratio follows $\frac{\alpha_1^{(2)}(Kk)}{\alpha_1^{(1)}(k)} = K^{2.5}$. This can be inferred from the definition of time scale $t_0 = \sqrt{\frac{m_0 l_0^2}{k_BT}} \sim l_0^{2.5}, (m_0 \sim l_0^3) $.  The numerical results in Figure \ref{Tgrowth4} also confirm this analysis.

The analysis and numerical results show that the growth rates can be positive in some intermediate wave number regime depicted in Figure \ref{Tgrowth4}, instead of near the zero wave number range. In this case, the growth rate of the Full model is the smallest while the EPNP model's is the highest. From the linear stability analysis, we notice that this instability is associated with a large interaction parameter $\xi_{12}$, a consequence of the finite size effect. A positive $\xi_{12}$ means that the interaction between different species due to their steric effects is repulsive. The analysis and numerical results tell us that the intermediate wave instability appears when the repulsive effect is sufficiently strong in the three models. This also can be obtained from the interaction free energy density $g$. The repulsive interaction due to the finite size effect is represented by $\frac{k_BT}{2}(\xi_{11}n_1^2 + 2\xi_{12}n_1n_2 + \xi_{22}n_2^2)$, which can be rewritten as  $\frac{k_BT}{2}((\sqrt{\xi_{11}}n_1 + \frac{\xi_{12}}{\sqrt{\xi_{11}}}n_2)^2 + (\xi_{22} - \frac{\xi_{12}^2}{\xi_{11}})n_2^2)$.
When $\xi_{12}$ is sufficiently large, $\xi_{11}\xi_{22} - \xi_{12}^2 < 0$, this quadratic form is hyperbolic type without lower bound.
In the next nonlinear simulations, we only consider the cases $\xi_{11}\xi_{22} - \xi_{12}^2>0$, with out the intermediate wave instability.

\subsubsection{Discussion on the finite size effect}
\noindent \indent
The hard sphere repulsion characterizes the finite-size effect of ions, witch keeps ions apart. The free energy density due to the finite-size effect is
\ben\bea{l}
\int K({\bf x}-{\bf y}) G(\{n_i\}_{i=1}^N({\bf x}), \{ n_i\}_{i=1}^N({\bf y}))d {\bf y} = \int \sum_{i=1}^N\sum_{j=1}^N\frac{\varepsilon_{ij}}{2}\frac{(a_i+a_j)^{12}}{|{\bf x}-{\bf y}|^{12}}n_i({\bf x})n_j({\bf y})d{\bf y},
\eea\een
where $a_i$ and $a_j$ are the radii of ion $i$ and $j$, and $\varepsilon_{ij}$ is the energy coupling constant between ion $i$ and $j$.
Thus, in the free energy function, we have the convolution integral with the following form
\ben\bea{l}
\int\int \frac{1}{|{\bf x}-{\bf y}|^{12}}n_i({\bf x})n_j({\bf y})d{\bf y}d{\bf x}.
\eea\een
We can approximate the above convolution integral by truncating the kernel $\frac{1}{|{\bf x}-{\bf y}|^{12}}$ with the cutoff length $\delta$. As discussed in the paper \cite{Horng2012}, when the cutoff length $\delta$ goes to zero, this convolution integral can be approximated by the integral
\ben\bea{l}
S_{\delta}\int n_i({\bf x})n_j({\bf y})d{\bf x},
\eea\een
with $S_{\delta} \approx \delta^{-12+d}$, where $d$ is the dimension.
The free energy density due to the finite-size effect can be written as
\ben\bea{l}
\sum_{i,j=1}^N \frac{\varepsilon_{ij}}{2}(a_i+a_j)^{12} S_{\delta} n_i({\bf x})n_j({\bf x}) = k_BT \sum_{i,j=1}^N \frac{\xi_{ij}}{2} n_in_j,
\eea\een
with $\xi_{ij} = \frac{1}{k_BT}\varepsilon_{ij}(a_i+a_j)^{12} S_{\delta}$. We add the conformational entropy in terms of the  derivative form to compensate for the approximation error, then the energy density for the finite-size effect is approximated by
\ben\bea{l}
g=k_BT[\sum_{i,j=1}^N \frac{\xi_{ij}}{2} n_i n_j+\sum_{i=1}^N \frac{\gamma_{i}}{2}\|\nabla n_i\|^2],
\eea\een
where $\gamma_{i}$ is a small parameter, witch can be zero. In the paper \cite{Horng2012},
the following $\varepsilon_{ij}$ values for the cross hard-sphere potential terms for some familiar ions ($Na^+,Cl^-,Ca^{2+}$) are used:
\ben\bea{l}
\varepsilon_{Na,Na}:\varepsilon_{Cl,Cl}:\varepsilon_{Ca,Ca}:\varepsilon_{Na,Cl}:
\varepsilon_{Na,Ca}:\varepsilon_{Cl,Ca} = 1:1:1:0.955:1:0.961.
\eea\een
Also in the paper \cite{Horng2012}, the ratios of the interaction coefficients $\xi_{ij}$ are given for some familiar ions ($Na^+,Cl^-,Ca^{2+}$) as follows
\ben\bea{l}
\xi_{Na,Na}:\xi_{Cl,Cl}:\xi_{Ca,Ca}:\xi_{Na,Cl}:\xi_{Na,Ca}:\xi_{Cl,Ca} = 1:2280:1.64:42.2:0.642:50.4.
\eea\een
It is easy to verify that $\xi_{11}\xi_{22} - \xi_{12}^2>0$ for two of the three ions. For the familiar ions, the interaction coefficients $\xi_{ij}$ are in the stable regime. That is the reason we only consider the stable cases in the nonlinear simulations next.

\subsection{Nonlinear dynamics}

\noindent \indent
We next explore nonlinear dynamics of the models in the linearly stable regime. We use the characteristic length scale $l_0 = 1nm$ and set the domain as $x \in [0,10]$. The values of the interaction parameters are chosen as $ \xi_{11}=\xi_{22}=1, \xi_{12} = 0.8$, satisfying $\xi_{11}\xi_{22} - \xi_{12}^2>0$. We also set diffusion coefficients $\gamma_1 = \gamma_2 = 0$.
The boundary conditions for the number densities $n_1, n_2$ are no-flux boundary conditions (\ref{nof}); 
for the velocity ${\bf v}_1$, the boundary conditions are set at  ${\bf v}_1|_{x = 0,10} = 0$, and for the electric potential $\Phi$, they are set at  $\Phi|_{x = 0} = 0, \Phi|_{x = 10} = \Phi_0$, where $\Phi_0$ is the electric potential at the right boundary $x=10$. We set $\Phi_0 = 1$ in the following simulations. The initial conditions are given by
$ n_1 = n_2 = n^0 = 1, \ {\bf v}_1 = 0, \ p =0,\  \Phi = \Phi_0 x/10$. The given external electric potential is $\Phi_e = \Phi_0 x/10$.
The dimensionless mobilities are given as $\lambda_1 = \lambda_2 = 0.02$.  We compute the ionic number densities using the Full model, the EPNP model and the CPNP model, respectively.

\begin{figure}[!ht]
\centering
\subfigure[Number density]
{\includegraphics[width=0.24\textwidth]{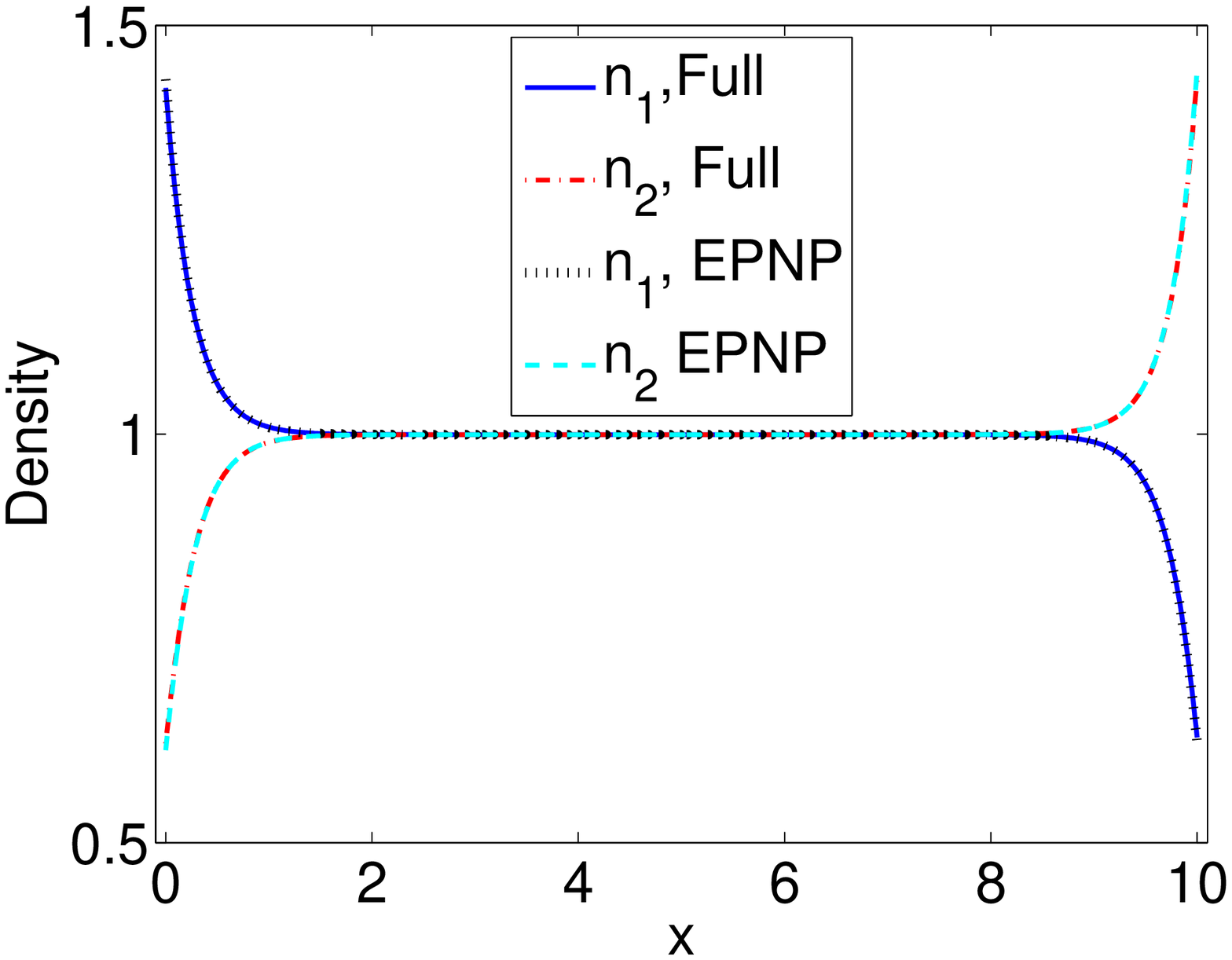}}
\subfigure[Differences between the models]
{\includegraphics[width=0.24\textwidth]{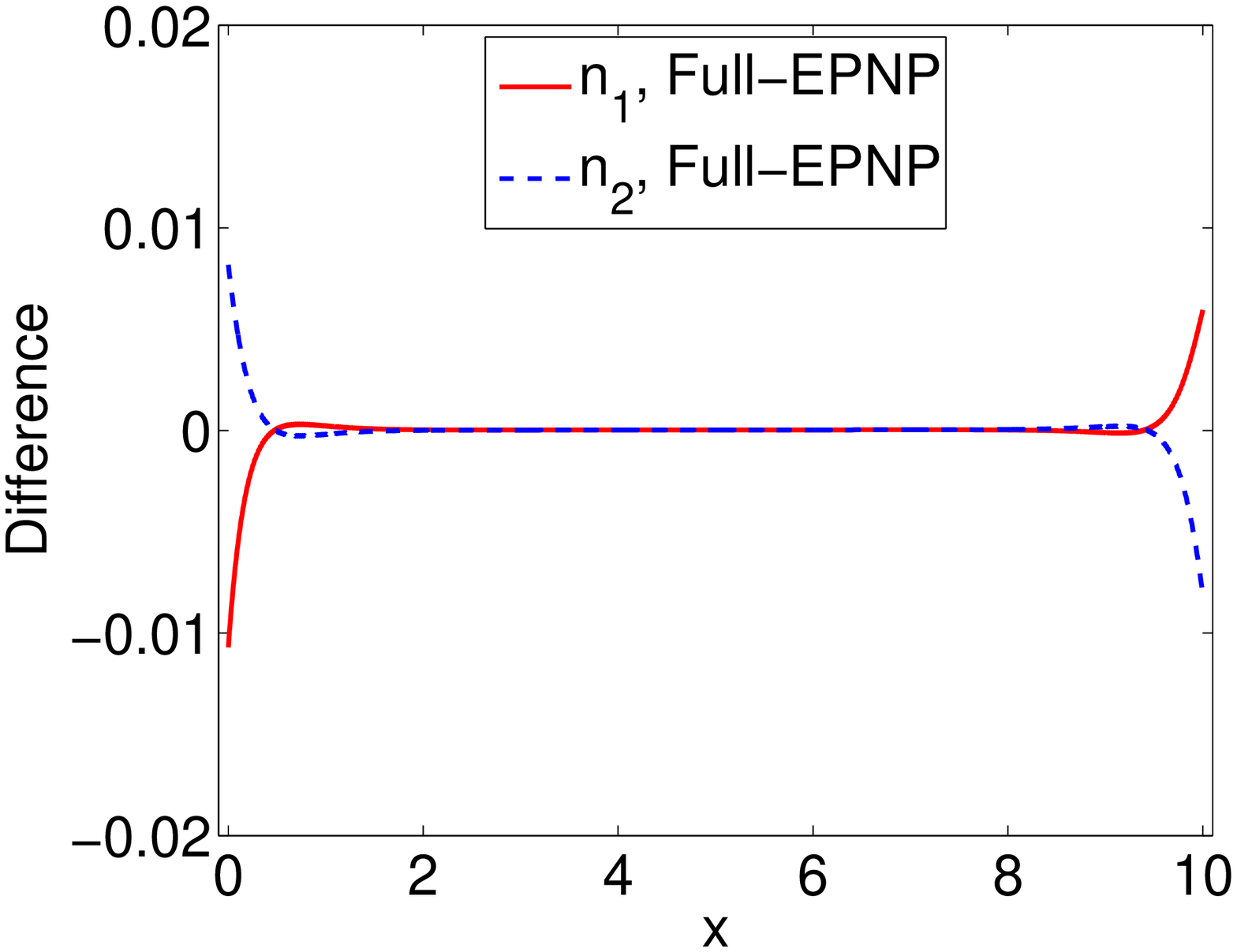}}
\subfigure[Electric potential $\Phi$]
{\includegraphics[width=0.24\textwidth]{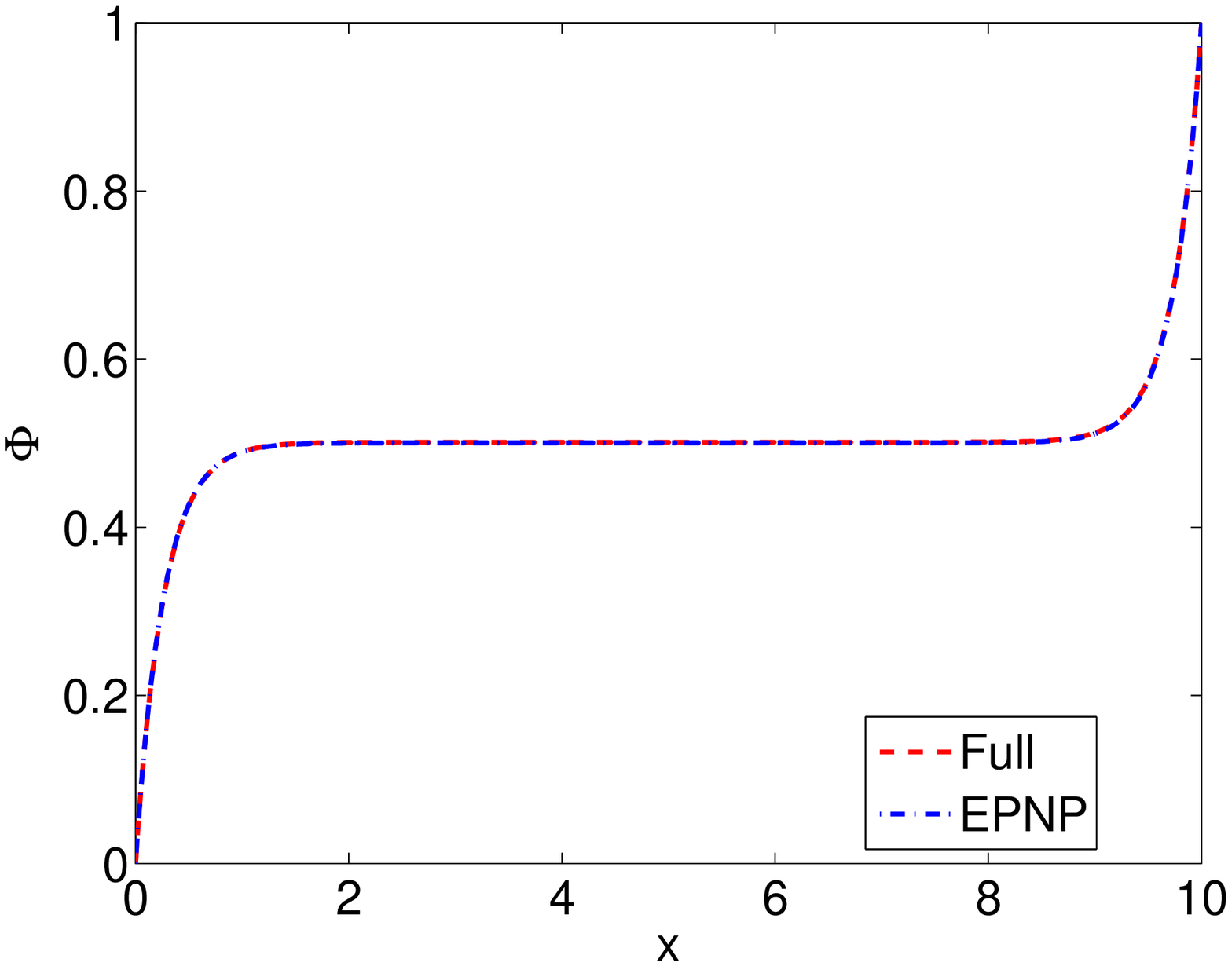}}
\subfigure[Energy $F(t)-F(0)$]
{\includegraphics[width=0.24\textwidth]{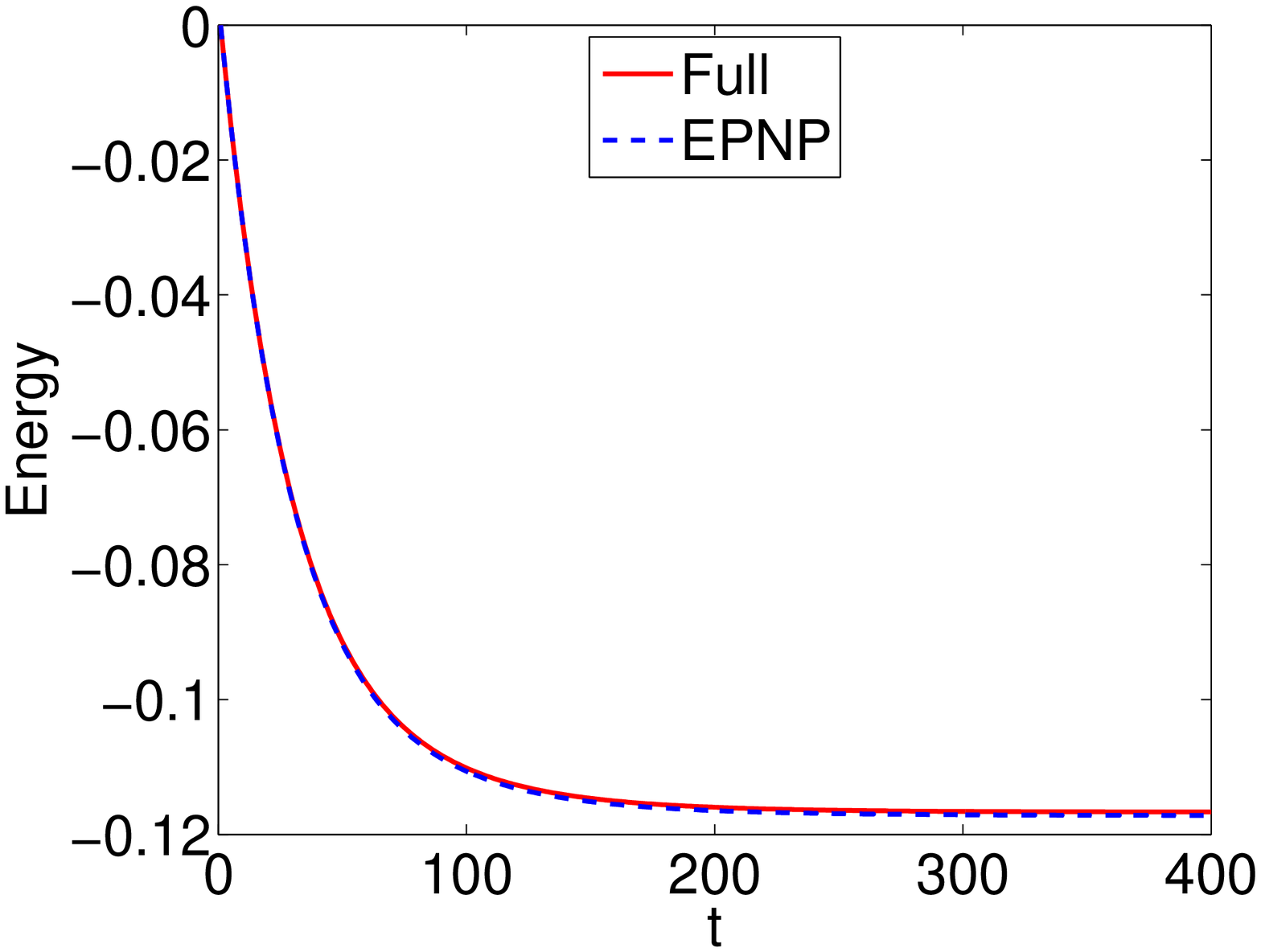}}
\caption{Steady states of the ionic densities and the electric potential of the Full and the EPNP models with $\xi_{11} = \xi_{22} = 1, \xi_{12} = 0.8$ in the stable regime, respectively. The differences appear near the boundary with absolute maximum difference $0.011$. The curve of energy difference $F(t)-F(0)$ is plotted with respect to time. The total free energy $F(t)$ decays to a constant when the final steady state is obtained. }
\label{fe}
\end{figure}

Figures \ref{fe} depicts the final steady states of the Full model and the EPNP model, and the difference between them, where $\xi_{11} = \xi_{22} = 1, \xi_{12} = 0.8$ in the stable regime. The states of the number densities are almost identical in the middle of the domain, while the visible differences appear near the two boundaries. Because the electric potential is positive at the right boundary and zero at the left boundary, some negative ions gather at the right side while positive ions gather at the left side due to the Coulomb force, forming two visible boundary layers.

As shown in Figure \ref{fe}, the density differences between the two models are about $O(1)\times 10^{-2}$ near the boundaries. As a conclusion, the compressibility of the flow in the full model plays relatively important role, it impacts the aggregation effect of the ions near the boundaries.

\begin{figure}[!ht]
\centering
\subfigure[$\rho_3:\rho_1:\rho_2 = 1:0.5:2$]
{\includegraphics[width=0.325\textwidth]{d_fe.eps}}
\subfigure[$\rho_3:\rho_1:\rho_2 = 1:0.25:4$]
{\includegraphics[width=0.325\textwidth]{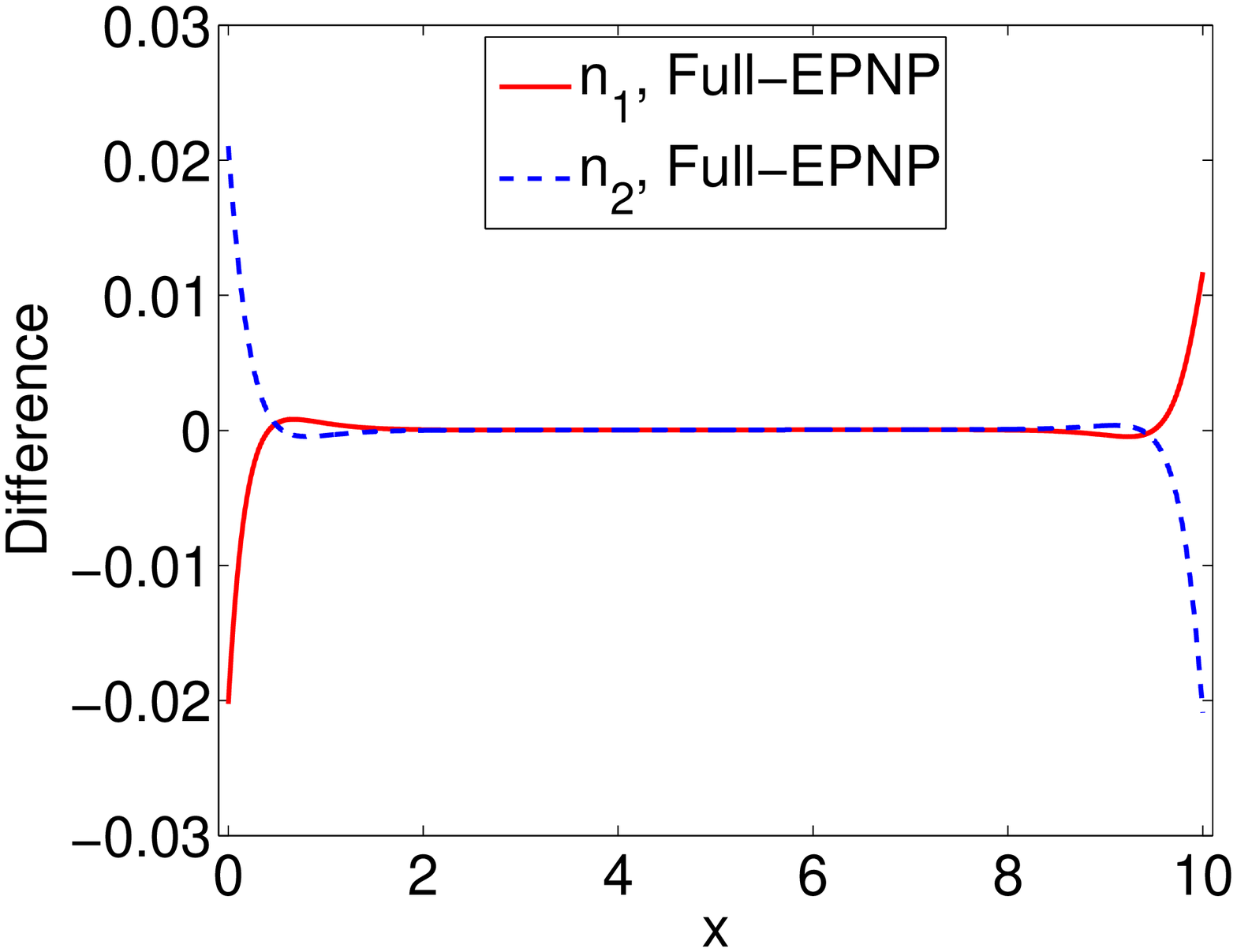}}
\subfigure[$\rho_3:\rho_1:\rho_2 = 1:0.125:8$]
{\includegraphics[width=0.325\textwidth]{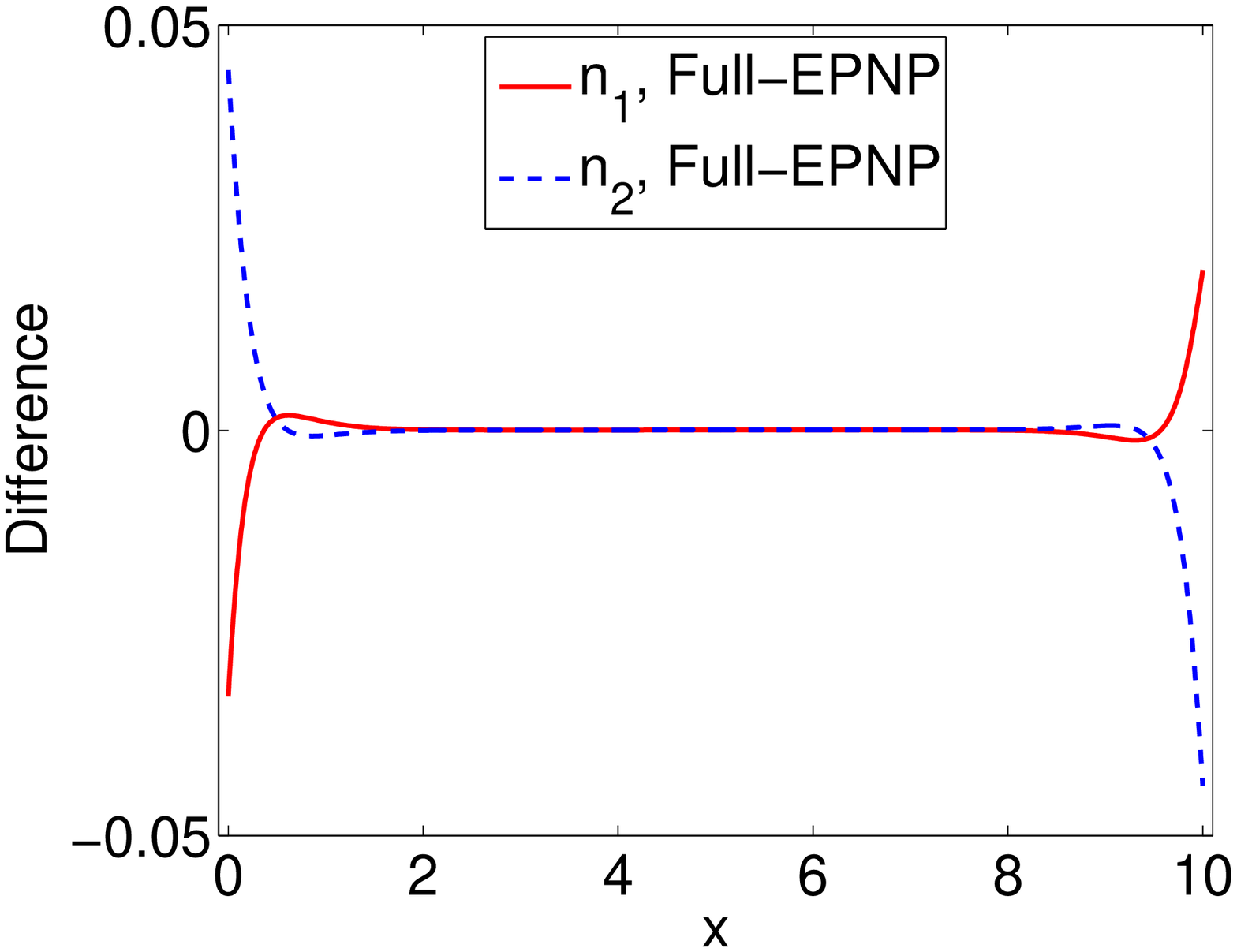}}
\caption{Differences between the Full and the EPNP model with $\xi_{11} = \xi_{22} = 1, \xi_{12} = 0.8$ in the stable regime. (a) The density ratio is $\rho_3:\rho_1:\rho_2 = 1:0.5:2$ and $R_1 = 0.025, R_2 = -0.025$, the absolute maximum difference is about $0.011$; (b) the density ratio is $\rho_3:\rho_1:\rho_2 = 1:0.25:4$ and $R_1 = 0.0375, R_2 = -0.075$, the absolute maximum difference is about $0.022$; (c) the density ratio is $\rho_3:\rho_1:\rho_2 = 1:0.125:8$ and $R_1 = 0.04375, R_2 = -0.175$, the absolute maximum difference is about $0.044$. }
\label{dfe_5}
\end{figure}

In the above example, the density ratio is chosen as $\rho_3:\rho_1:\rho_2 = 1:0.5:2$. By halving density $\rho_1$ and doubling  density $\rho_2$ to increase the density differences, we reset the density ratio as $\rho_3:\rho_1:\rho_2 = 1:0.25:4$ and $\rho_3:\rho_1:\rho_2 = 1:0.125:8$, while maintaining the volume ratio unchanged at $v_3:v_1:v_2=1:2:1$, then the dimensionless parameters are $R_1 = 0.0375, R_2 = -0.075$ and $R_1 = 0.04375, R_2 = -0.175$, respectively. In these cases, the size differences of the three components become larger. As shown in Figure \ref{dfe_5}, the differences between the Full model and the EPNP model become larger as the size differences  become larger. When we halve the density $\rho_1$ and double the density $\rho_2$, the absolute maximum difference between the Full model and the EPNP model is almost doubled.  As a result, when the size differences between the components are enlarged, the parameters $R_1, R_2$ are no longer small so that the compressibility of the flow can no longer be neglected.

\begin{figure}[!ht]
\centering
\subfigure[Number density]
{\includegraphics[width=0.325\textwidth]{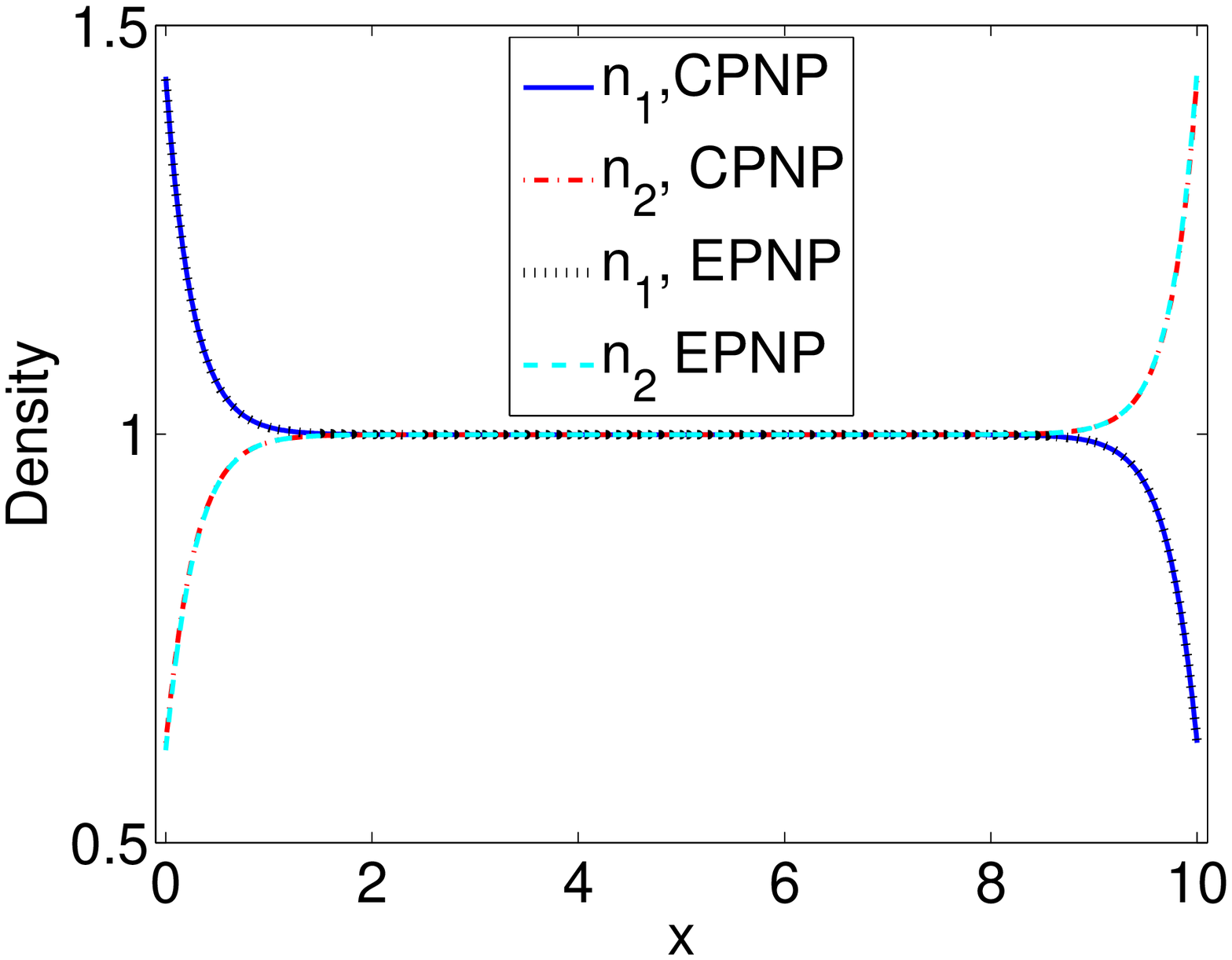}}
\subfigure[Difference between the models]
{\includegraphics[width=0.325\textwidth]{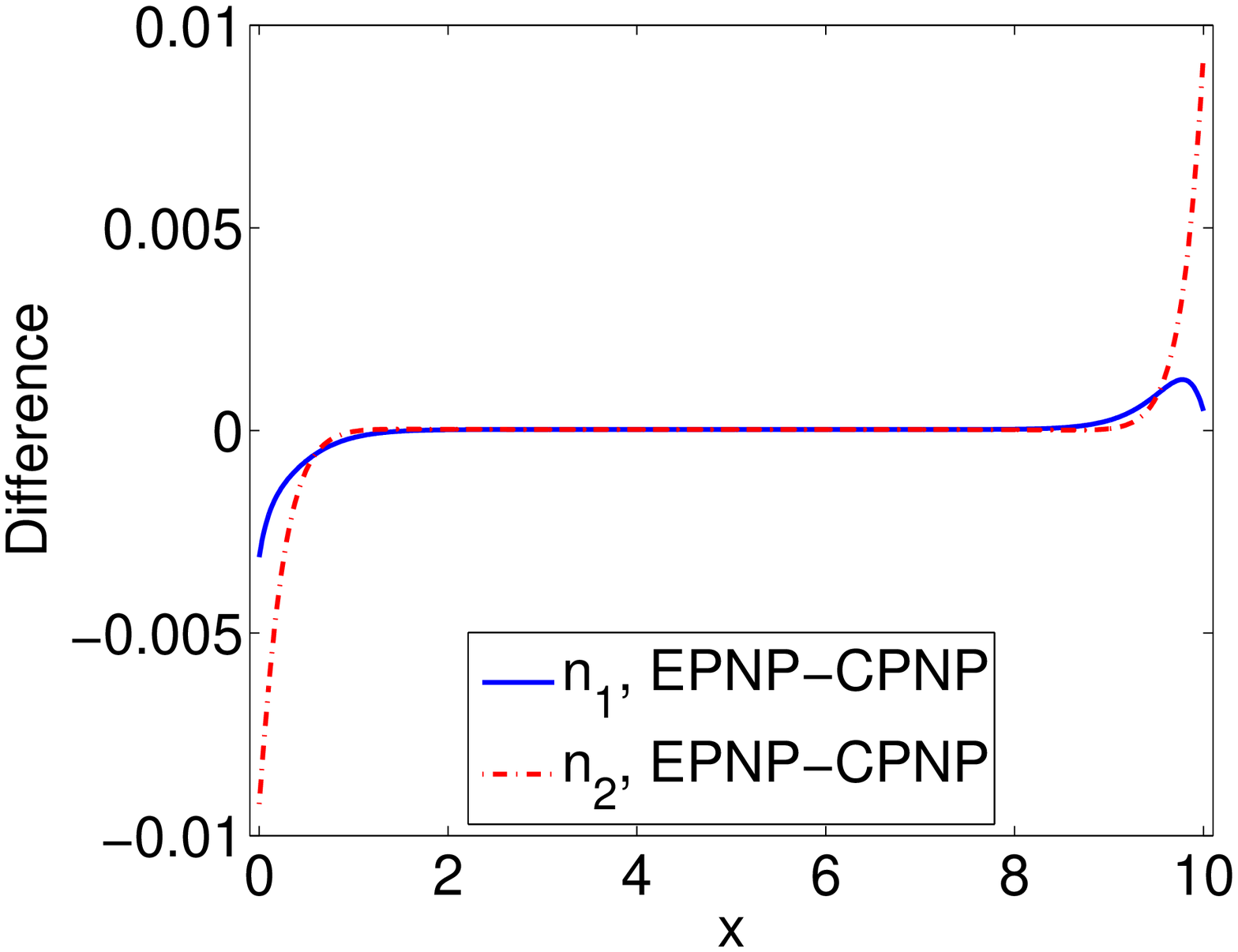}}
\subfigure[Electric potential $\Phi$]
{\includegraphics[width=0.325\textwidth]{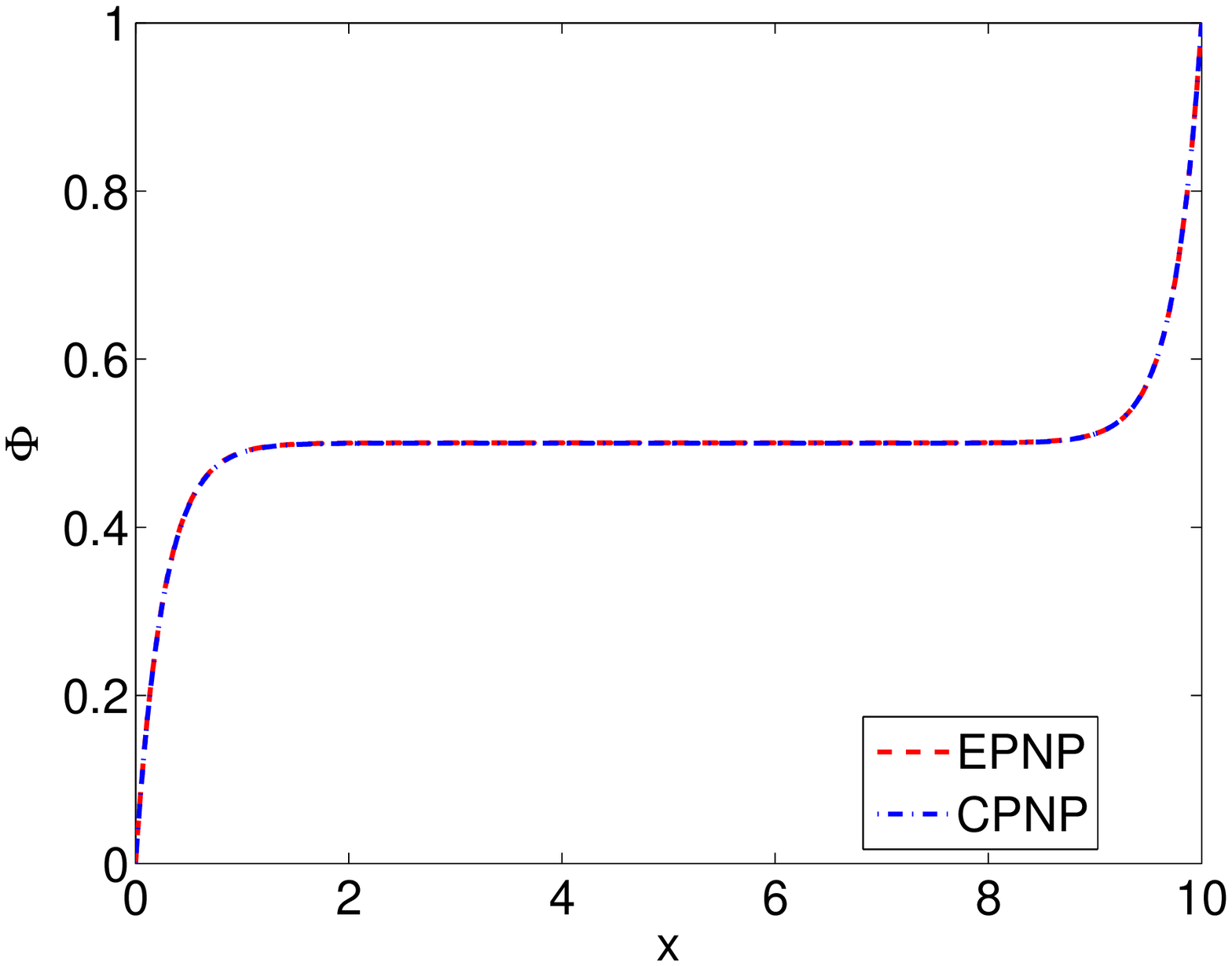}}
\caption{Steady states of the ionic densities and electric potential of the EPNP and the CPNP model with $\xi_{11} = \xi_{22} = 1, \xi_{12} = 0.8$ in the stable regime. The difference appears at the boundary layers near both ends of the domain. }
\label{ec}
\end{figure}

We also compare the EPNP and the CPNP model in the stable regime with $\xi_{11} = \xi_{22} = 1, \xi_{12} = 0.8$ in Figure \ref{ec}. The density ratio $\rho_3:\rho_1:\rho_2 = 1:0.5:2$ is used. The values of the parameters are set at $r_1^m=1, r_2^m = 2$. The number density differences between the two models are about $O(1)\times 10^{-2}$ near the boundaries. The differences near the boundaries in $n_2$ are bigger than that in $n_1$ . The reason is that in the CPNP model, the term $r_i^m\mu_3$ is dropped in the $n_i$ transport equation. In this example, $r_2^m > r_1^m$, so the differences near the boundaries of $n_2$ are bigger.
Consequently, the solvent's chemical potential $\mu_3$ in the EPNP model plays a relatively important role, it impacts the aggregation effect of the ions near the boundaries.

\begin{figure}[!ht]
\centering
\subfigure[Number density]
{\includegraphics[width=0.325\textwidth]{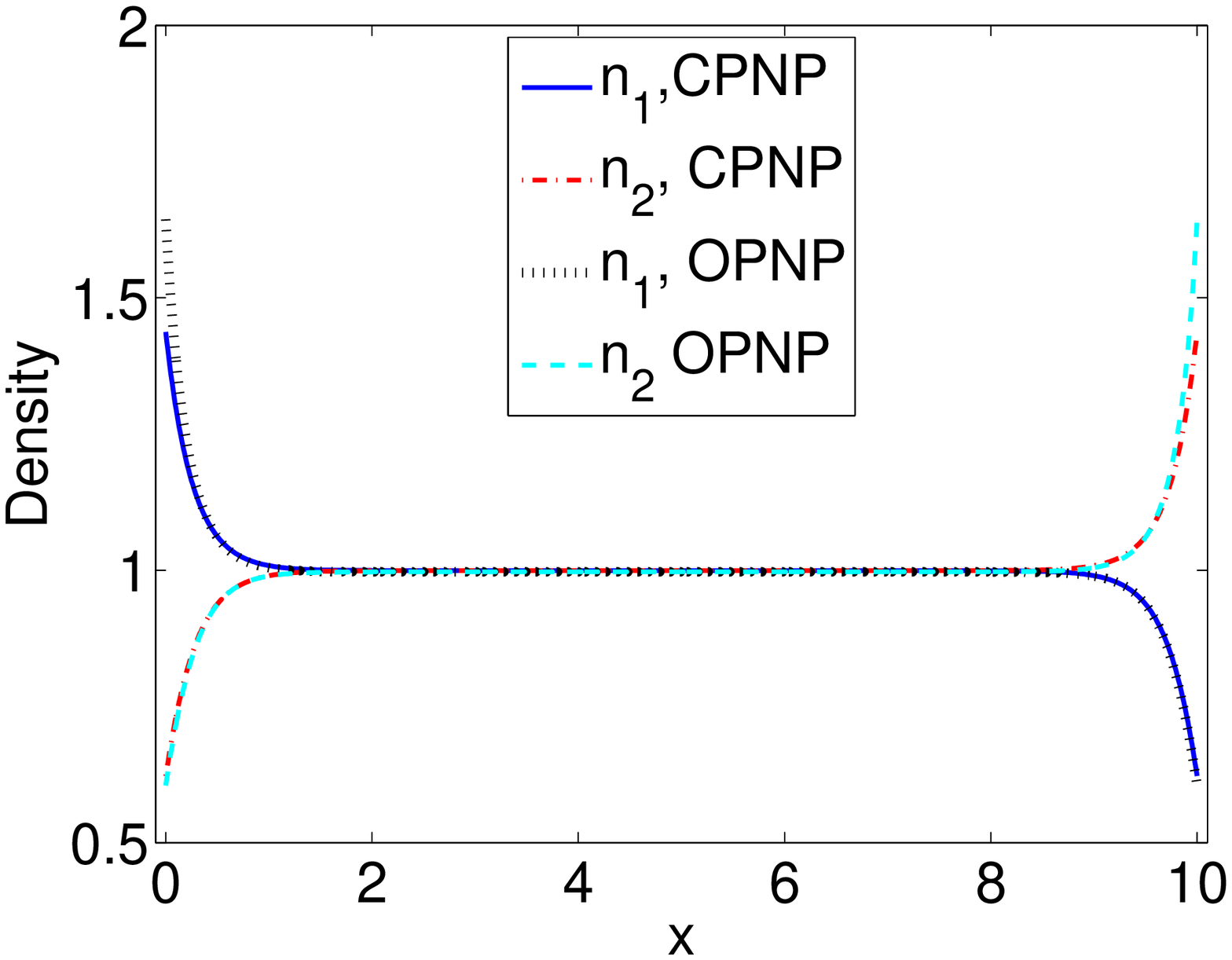}}
\subfigure[Difference between the models]
{\includegraphics[width=0.325\textwidth]{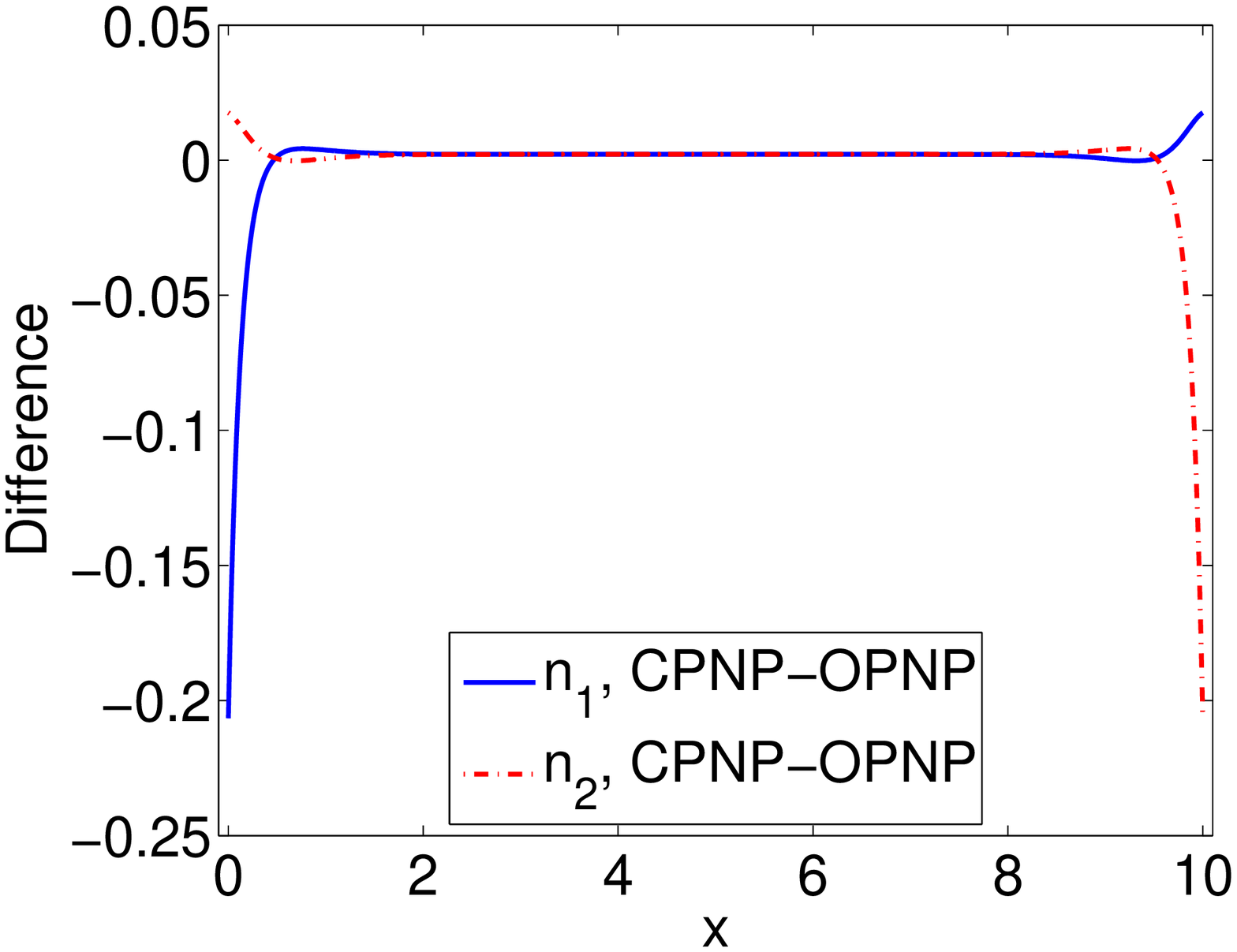}}
\subfigure[Electric potential $\Phi$]
{\includegraphics[width=0.325\textwidth]{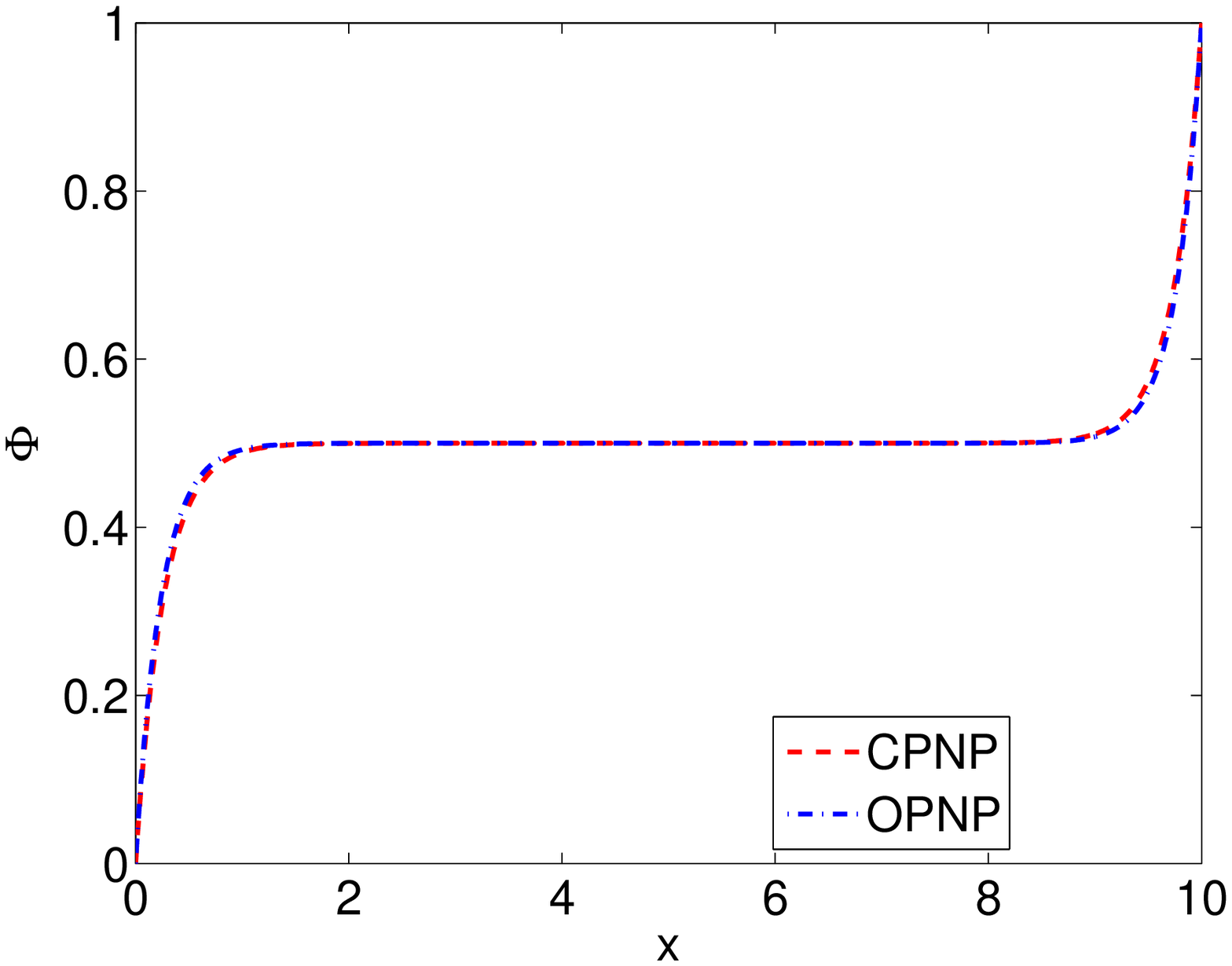}}
\caption{Steady states of ionic densities and the electric potential of the CPNP and the OPNP model. In the CPNP model $\xi_{11} = \xi_{22} = 1, \xi_{12} = 0.8$ in the stable regime. The difference appears at the boundary layers near both ends of the domain. }
\label{co}
\end{figure}

Next, we consider ionic concentrations without the finite size effect and compare them with ionic concentrations with the finite size effect using the classical PNP model. In the following, the OPNP means the classical PNP model without finite size effects (i.e., $\xi_{11}=\xi_{22}=\xi_{12} = 0$  and $\gamma_1=\gamma_2 =0$.) The differences also appear in the areas near the two boundaries. As shown in Figures \ref{co}, the differences of the ionic density can reach up to $O(1)\times 10^{-1}$. The finite size effect plays an important role in the system, it impacts the aggregation effect of the ions near the boundaries, as studied in the papers \cite{Hyon&E&L2011,Hyon&E&L2014,Lin&E2015,Horng2012}.

Based on our numerical investigations and the linear analysis, we conclude that the 1D steady states of the number densities are nearly identical in the middle of the domain in all three models in the stable regime. The differences lie in the areas near the boundaries. The compressibility of the flow, the chemical potential of the solvent and the finite size effect are three main reasons that lead to the differences. So, our quasi-incompressible model (the full model) seems to be more reasonable  because the  mass and momentum conservation laws are preserved in the model while the other models don't respect the two fundamental physical conservation laws.

Further investigations in higher dimensions is necessary to evaluate the difference among the models, which will be conducted in a sequel.

\section{Conclusion}

\noindent \indent
We have developed systematically a set of quasi-incompressible theories for ionic fluids of multiple species that respect not only momentum conservation but also mass conservation at the presence of the ionic species. The previous PNP type models are approximations of the more fundamental theories when densities of different ionic species are distinct. In these theories, we consider the entropic contribution from each ionic species together with the ion-ion interaction due to the finite size effect. The limiting cases include the extended PNP, the classical PNP with the finite size effect, and the classical PNP model without the finite size effect. At the length scale larger than hundreds of nanometers, all models agree with the classical PNP model very well. At the length scale in a few nanometers, the models can predict quite different stability behavior for homogeneous equilibrium states.
In nonlinear dynamics, the ionic number densities are nearly identical in the middle of the domain, but the differences lie in the areas near the boundaries. Apparently, three main factors in the compressibility of the flow, the chemical potential of the solvent and the finite size effect of the ions can lead to the discrepancy in model predictions. We tend to believe that the new model is more accurate since it obeys the two fundamental physical conservation laws in mass and linear momentum while the others don't.

\bigskip
\noindent {\bf Acknowledgment}
\vskip 12 pt Xiaogang Yang's work is supported by the Scientific Research Fund of Wuhan Institute of Technology through Grants K201741; Jun Li's work is partially supported by NSF of China through a grant (NSFC-11301287); Qi Wang is partially supported by  NSF through awards DMS-1200487 and DMS-1517347 as well as a grant from NSFC \# 11571032 and \# 91630207.

\section{Appendix}

The linearized eigenvalue problem for the Full model is formulated as follows,
\begin{equation}
\left(
\left(
\begin{array}{ccccc}
 0 & 0 & 0 & R_1\alpha & R_2\alpha \\
 0 & 0 & 0 & 0 & 0 \\
 0 & 0 & \rho_k\alpha & 0 & 0 \\
 0 & 0 & 0 & \alpha & 0  \\
 0 & 0 & 0 & 0 & \alpha  \\
\end{array}
\right)
+
\left(
\begin{array}{ccccc}
 0 & 0 & -ik\rho_k & 0 & 0 \\
 0 & -k^2 & 0 & \frac{1}{\varepsilon} & -\frac{1}{\varepsilon} \\
 ik & 0 & (\frac{4}{3}\eta+\nu) k^2 & B_1 &  B_2\\
 R_1\lambda_1n^0k^2 & \lambda_1n^0k^2  & ikn^0 & C_1 & C_2\\
 R_2\lambda_2n^0k^2 & -\lambda_2n^0k^2 & ikn^0 & D_1 & D_2 \\
\end{array}
\right)
\right)
\left(
\begin{array}{c}
p^0  \\
\Phi^0 \\
 {\bf v}_1^0\\
 n_1^0\\
 n_2^0\\
\end{array}
\right)
= 0 ,
\label{coefficient}
\end{equation}
where the parameter values  are given by $n_3^0 = r_0^v-r^v_1 n^0 - r^v_2 n^0$ and $\rho_k = 1 -R_1 n^0- R_2n^0$. The other components in the matrix are defined as follows
\ben\bea{l}
B_1 =  -ik r_1^v + ik n^0(\frac{1}{N_1n^0} + \xi_{11} + \xi_{12}) + ik^3n^0\gamma_1,\\
B_2 =  -ik r_2^v + ik n^0(\frac{1}{N_2n^0} + \xi_{22} + \xi_{12}) + ik^3n^0\gamma_2,\\
C_1 = \lambda_1 n^0 r_1^m r_1^v \frac{k^2}{n_3^0} + \lambda_1n^0(\frac{1}{N_1n^0} + \xi_{11})k^2 + \lambda_1n^0\gamma_1k^4,\\
C_2 = \lambda_1 n^0 r_1^m r_2^v \frac{k^2}{n_3^0} + \lambda_1n^0(\xi_{12})k^2,\\
D_1 = \lambda_2 n^0 r_2^m r_1^v \frac{k^2}{n_3^0} + \lambda_2n^0(\xi_{12})k^2,\\
D_2 = \lambda_2 n^0 r_2^m r_2^v \frac{k^2}{n_3^0} + \lambda_2n^0(\frac{1}{N_2n^0} + \xi_{22})k^2 + \lambda_2n^0\gamma_2k^4.
\eea\een
Although the coefficient matrix is $5 \times 5$, the characteristic polynomial of the coefficient matrix is a third order polynomial of growth rate $\alpha$, which yields three independent eigen-modes. Using an asymptotic analysis at small wave numbers $|k| \ll 1$, the three asymptotic growth rates are obtained asymptotically:
\ben\bea{l}
\alpha_1 = T_1 k^2 + O(k^4),\\
\\
\alpha_{2,3} = - \frac{T_2 \pm \sqrt{T_2^2 + T_3}}{T_4} + O(k^2) ,
\eea\een
where
\ben\bea{l}
T_1 = -\frac{\lambda_1\lambda_2 n^0}{(\lambda_1+\lambda_2)n_3^0}[{\rho_k^2 n_3^0}(\frac{1}{N_1 n^0} + \frac{1}{N_2 n^0} + \xi_{11}+\xi_{22}+ 2\xi_{12}) +  \rho_k (r_1^v+r_2^v)( (r_1^m+r_2^m)+ (R_1+R_2)n_3^0)] ,\\
\\
T_2 = \lambda_1\lambda_2\rho_k (n^0)^2(R_1+R_2)^2  + \varepsilon ,\\
\\
T_3 = -4 (n^0)^2\rho_k \varepsilon(\lambda_1 + \lambda_2)(\lambda_1 R_1^2 + \lambda_2 R_2^2),\\
\\
T_4 = 2n^0 \rho_k \varepsilon(\lambda_1 R_1^2 + \lambda_2 R_2^2).
\eea\een
Notice that $\alpha_{2,3}<0$ for small $k$ due to $T_3 < 0$. The eigenvalue $\alpha_1 > 0$ when $T_1 > 0$,
\ben\bea{l}
\frac{1}{N_1 n^0} + \frac{1}{N_2 n^0} + \xi_{11}+\xi_{22}+ 2\xi_{12} <  -\frac{r_1^v+r_2^v}{\rho_k n_3^0}( (r_1^m+r_2^m)+ (R_1+R_2)n_3^0).
\label{unst}
\eea\een
This is the instability condition for long waves for the Full model.
It follows from eqn (\ref{solv}) that  $ (r_1^m+r_2^m)+ (R_1+R_2)n_3^0>0$. So,  the instability can incur only when $\xi_{11}+\xi_{22}+ 2\xi_{12}$ is negative enough. But, $\xi_{ij}>0$ in the model. So this mode of instability is absent from the full model.

For the EPNP model, only $n_1, n_2, \Phi$ are coupled, the eigenvalue problem is given by
\begin{equation}
\left(
\left(
\begin{array}{ccc}
 0 & 0 & 0 \\
 0 & \alpha & 0  \\
 0 & 0 & \alpha  \\
\end{array}
\right)
+
\left(
\begin{array}{ccc}
 -k^2 & \frac{1}{\varepsilon} & -\frac{1}{\varepsilon} \\
\lambda_1n^0k^2   & C_1 & C_2\\
-\lambda_2n^0k^2 & D_1 & D_2 \\
\end{array}
\right)
\right)
\left(
\begin{array}{c}
\Phi^0 \\
 n_1^0\\
 n_2^0\\
\end{array}
\right)
= 0 ,
\label{coefficient2}
\end{equation}
where
\ben\bea{l}
C_1 = \lambda_1 n^0 r_1^m r_1^v \frac{k^2}{n_3^0} + \lambda_1 n^0(\frac{1}{N_1 n^0} + \xi_{11})k^2 + \lambda_1 n^0 \gamma_1 k^4,\quad C_2 = \lambda_1 n^0 r_1^m r_2^v \frac{k^2}{n_3^0}+ \lambda_1 n^0 \xi_{12} k^2,\\
\\
D_1 = \lambda_2 n^0 r_2^m r_1^v \frac{k^2}{n_3^0} + \lambda_2 n^0 \xi_{12} k^2, \quad D_2 = \lambda_2 n^0 r_2^m r_2^v \frac{k^2}{n_3^0} + \lambda_2 n^0(\frac{1}{N_2 n^0} + \xi_{22})k^2 + \lambda_2 n^0 \gamma_2 k^4.
\eea\een
Eliminating $\Phi^0$, the system  reduces to
\begin{equation}
\left(
\left(
\begin{array}{cc}
 \alpha & 0  \\
 0 & \alpha  \\
\end{array}
\right)
+
\left(
\begin{array}{cc}
 C_1+ \lambda_1 n^0 \frac{1}{\varepsilon} & C_2 - \lambda_1 n^0 \frac{1}{\varepsilon}\\
 D_1- \lambda_2 n^0 \frac{1}{\varepsilon} & D_2 + \lambda_2 n^0 \frac{1}{\varepsilon}\\
\end{array}
\right)
\right)
\left(
\begin{array}{c}
 n_1^0\\
 n_2^0\\
\end{array}
\right)
= 0 ,
\label{coefficient3}
\end{equation}
The characteristic polynomial is quadratic and given by
\ben\bea{l}
\alpha^2 + [C_1+D_2 + (\lambda_1+\lambda_2)n^0\frac{1}{\varepsilon}]\alpha + A = 0,\\
\\
A = C_1D_2 - C_2D_1 + [\lambda_2(C_1+C_2)+\lambda_1(D_1+D_2)]n^0\frac{1}{\varepsilon}.
\eea\een
The two growth rates are given by
\ben
\bea{l}
\alpha_1 = -2A\left [(C_1+D_2 + (\lambda_1+\lambda_2)n^0\frac{1}{\varepsilon})+\sqrt{(C_1+D_2 + (\lambda_1+\lambda_2)n^0\frac{1}{\varepsilon})^2-4A}\right ]^{-1},\\
\\
\alpha_2 = -\frac{1}{2}\left [(C_1+D_2 + (\lambda_1+\lambda_2)n^0\frac{1}{\varepsilon})+\sqrt{(C_1+D_2 + (\lambda_1+\lambda_2)n^0\frac{1}{\varepsilon})^2-4A} \right].
\eea
\een
Notice that $C_1 \geq 0, D_2 \geq 0$. So, $Re(\alpha_2)<0$ and $Re(\alpha_1)$ can be positive only if $A<0$.

\end{document}